\documentclass[11pt]{article}
\usepackage{epsfig} 
\begin{document}

\begin{titlepage}
\begin{center}

Draft: \today	
{}~             
{}~		

\vskip0.7in 
{\Large\bf Domain walls out of equilibrium}

\vskip.3in

S.M. Alamoudi$^{(a)}$\footnote{Email: {\tt smast15@vms.cis.pitt.edu}},
D. Boyanovsky$^{(a)}$\footnote{Email: {\tt boyan@vms.cis.pitt.edu}} and 
F.I. Takakura$^{(a,b)}$\footnote{Email: {\tt takakura@fisica.ufjf.br}}\\ 

\bigskip

{\it (a) Dept.\ of Physics and Astronomy, University of Pittsburgh, 
         Pittsburgh PA USA 15260}\\
{\it (b) Departamento de F\'{\i}sica, Instituto de Ci\^encias Exatas, \\
Universidade Federal de Juiz de Fora, Juiz de Fora, 36036-330, MG, Brasil}\\

\vskip.3in

\end{center}

\vskip.5in

\begin{abstract}
We study the non-equilibrium dynamics of  domain walls in real time
for $\phi^4$ and Sine Gordon models in $1+1$ dimensions in the dilute regime. 
The equation of motion for the collective
coordinate is obtained by integrating out the meson excitations around the
domain wall to one-loop order. The real-time non-equilibrium relaxation is studied  analytically and numerically to this order.  The
constant friction coefficient vanishes but there is dynamical friction and
relaxation caused by off-shell non-Markovian effects. The validity of a Markovian description
is studied in detail. The proper Langevin 
equation  is obtained to this order, the noise is Gaussian and additive but colored. We analyze the
classical and hard thermal loop contributions to the self-energy and noise
kernels and show that  at temperatures larger than the meson mass the hard
contributions are negligible and the finite temperature contribution to the dynamics is governed by the classical soft modes of the meson bath. The long time relaxational dynamics is
completely dominated by classical Landau damping  resulting in that  the corresponding time scales are not set
by the temperature but by the meson mass.  The noise correlation function and the dissipative kernel obey a generalized form of the Fluctuation-Dissipation relation.

\end{abstract}
\end{titlepage}

\setcounter{footnote}{0}


\section{INTRODUCTION AND MOTIVATION \label{secintro}}
Kinks and domain walls play a fundamental role in the equilibrium and non-equilibrium processes after phase transitions to broken symmetry states in
theories with scalar order parameters. In scalar field theories that undergo
a symmetry breaking phase transition the process of phase ordering proceeds
by the formation of domains of the ordered phase separated by domain walls. 
These domain walls are topological defects that separate regions in which
the order parameter is locally ordered and therefore locally the system is
in a broken symmetry ground state in each domain. Interest in the dynamics of
these topological excitations is interdisciplinary. In condensed matter systems
solitons (or kinks) are collective excitations in quasi-one dimensional charge density wave systems and conducting polymers\cite{schrieffer,yu,gruner}. In particle physics domain walls in the form of sphalerons\cite{kuzmin} have been
argued to play an important role in baryogenesis\cite{kaplan}, and in cosmology domain walls and other topological defects could be responsible for structure formation\cite{kolb,vilenkin}.   

The classical statistical mechanics of a gas of kinks in one spatial dimension
has been previously studied\cite{habib} and it was established that the kink 
density at a temperature T is approximately given by $n_K \approx e^{-M/T}$
with $M$ being the kink mass. Therefore a study of the dynamics of single
domain walls or kinks will be valid in the dilute regime $M>>T$ in which
the interaction between kinks can be ignored because the mean separation
between kinks is much larger than the typical width of a kink (of the
order of the zero temperature correlation length or inverse meson mass).

The focus of this article is to study the \textit{real time} dynamics of relaxation
of domain walls (kinks) in $1+1$ dimensions via the interaction between the domain wall and the meson fluctuations in model field theories.
In particular we study in detail scalar $\phi^4$ and Sine-Gordon kinks in
the dilute regime in which $T << M$.  
 This problem is important in particle physics, condensed matter and cosmology. In particle physics dissipative processes on the dynamics of sphaleron configurations are important to establish corrections
to the proper transition rates. In quasi-one dimensional condensed matter systems
kinks and domain walls are responsible for important transport phenomena and therefore a study of the dissipative aspects will provide a deeper understanding of these phenomena. In cosmology the evolution of domain walls or dynamic of
interfaces determines the scales in which ordering of horizon sized regions occur.

Although a study in $1+1$ dimension may not be a proper realization of the
$3+1$ dimensional situations in particle physics and cosmology, it will at
least highlight important aspects of the dynamics that must be generalized to the proper situations. 

In condensed matter there is a considerable effort in understanding dissipative
aspects of solitons starting from a microscopic description\cite{rev}-\cite{neto}
in terms of Mori's formulation of linear response, and more recently
in terms of a system-bath formulation\cite{neto}.

Recently Khlebnikov\cite{khleb} has studied the velocity of a bubble wall in the case of a
non-degenerate scalar potentials. The bubble-wall velocity was related to the self-energy of 
the scalar field through the fluctuation-dissipation theorem. Assuming a trilinear coupling to
another massive field a local friction coefficient was extracted. Alternatively,
Arnold\cite{arnold} provided an equivalent result to the one obtained in Ref. \cite{khleb} 
at one loop using reflection and transmission coefficients for particles scattering off 
the bubble wall.

Our approach is rather different. It is tailored to obtain a real time description of the dissipative
processes and a consistent derivation of the Langevin equation in a weak coupling
 perturbative expansion. The main ingredient is the collective coordinate quantization of the kink that allows
to obtain the non-equilibrium generating functional for the collective coordinate
by integrating out the meson degrees of freedom, i.e. the fluctuations around the kink. 
The resulting Langevin equation allows an unambiguous identification
of the dissipative kernel and the noise correlation function that obey 
 a generalized fluctuation dissipation relation. The dissipative processes arise from the interaction between the collective coordinate and the orthogonal fluctuations around the kink, 
rather than from the coupling to other fields.

We provide analytic and numerical study of the solutions of the
equations of motion of the kink collective coordinate in lowest order (one loop) and establish that a Markovian approximation fails to describe the dynamics at large temperatures. Furthermore we analyze in detail
the high temperature low density regime in which $m<<T<<M$ with
$m$ being the meson mass  focusing on the classical and
hard thermal loop contribution to the dissipative kernel and the noise-noise correlation function. We argue that in lowest order in perturbation theory, the long time dynamics  is completely dominated by classical Landau damping.  

The main results of this article are the following
\begin{itemize}
\item A field-theoretical derivation of the real-time non-equilibrium equations of motion of 
the collective coordinate associated with translations of the domain wall and its solution in 
relevant cases for the sine-Gordon and $\phi^4$ potentials.

\item A detailed microscopic derivation of the non-equilibrium influence functional, the
quantum Langevin equation and the generalized fluctuation dissipation theorem to one-loop 
order.

\item A detailed analytic and numerical study of the relaxation in the one-loop approximation.
The Markovian approximation is compared to the ``exact'' dynamics in a wide range of temperature
and the high temperature and classical limits analyzed in detail. The long time dynamics to 
this order is analyzed both analytically and numerically.
\end{itemize}

To our knowledge these aspects of domain wall dynamics had not been studied previously.

Section \ref{seccoll} summarizes briefly the main concepts in collective coordinate quantization that are relevant for our study. In section \ref{secDMbath} we introduce the main
tools of non-equilibrium field theory to study the kink in a bath of mesons
in equilibrium and describe in general the relevant interactions, the
equation of motion of the collective coordinate a Markovian approximation and the Langevin equation. 
Section \ref{secmodels} analyzes in detail the $\phi^4$ and Sine-Gordon models for which 
a Markovian approximation is shown to fail at large temperatures. In section
\ref{secHTloops} we study the high temperature  but low density limit ($m<<T<<M$) and 
establish that the long time dynamics is dominated by classical Landau damping processes. 
In section \ref{hidim} we discuss higher order corrections to the results obtained within 
the one-loop approximation and we comment on generalization to higher dimensions. 
Section \ref{seccon} presents our conclusions. Several appendices are included for 
technical details, in particular appendix \ref{C} establishes the generalized
fluctuation-dissipation relation between the damping kernel and the noise-noise 
correlation function.

\section{COLLECTIVE COORDINATE QUANTIZATION \label{seccoll}}
To begin our study of the dynamics of kinks we focus on $1+1$ dimensional
quantum field theories described by Hamiltonians of
the form
\begin{equation}
H\,=\,\int dx\,\left\{\frac{\pi^2}{2}+\frac{1}{2}\left(\frac{d\phi}{dx}\right)^2+U(\phi)\right\}
\label{hs}
\end{equation}
in which the potential $U(\phi)$ admits degenerate, broken symmetry minima. 

A static kink is a solution of the time independent field equation 
\begin{equation}
-\frac{d^2\phi_s}{dx^2}+\frac{\partial U(\phi_s)}{\partial\phi}\,=\,0 \label{eqnm}
\end{equation}
with boundary conditions such that $  \phi_s(x \rightarrow \pm \infty)=
\phi_{\pm \infty}$ and$ U(\phi_{\pm \infty}) = 0$ \cite{c7}-\cite{c51}.
Translational invariance implies that such solution is of the form
$\phi_s(x-x_0)$ with $x_0$ an arbitrary translation chosen such that $\phi_s(0)=0$, therefore $x_0$ is identified with the position of the kink.

Lorentz invariance  results in that a kink moving with constant velocity is given by $\phi_s\left[
\frac{x-x_0-vt}{\sqrt{1-v^2}}\right]$\cite{c7}-\cite{lee}. The mass of the kink, i.e, the energy of a static kink is given by
\begin{equation}
M \equiv E[\phi_s] = \int {\rm d} x \left({{{\rm d} \phi_s}\over{{\rm d} x}}\right)^2.
\label{M}
\end{equation}

Quantization around the static kink solution implies writing
\begin{equation}
\hat{\phi}(x,t)\,=\,\phi_s(x - x_0)+\hat{\psi}(x - x_0;t). \label{ex}
\end{equation}
Where the fluctuation operator is expanded in terms of a complete set of harmonic modes around the kink 

\begin{equation}
\hat{\psi}(x-x_0;t)\,=\,\sum_{n}^\infty q_n(t)\, {\cal U}_n(x - x_0)
\label{fluk}
\end{equation}

\noindent where the mode functions  ${\cal U}_n(x - x_0)$  obey

\begin{equation}
\left[- {{\rm d}^2\over{{\rm d} x^2}} + \left.\frac{d^2U}{d\phi^2}\right|_{\phi_s} \right] 
{\cal U}_n(x - x_0) = \omega_n^2 \, {\cal U}_n(x - x_0) 
\label{hosc}
\end{equation}

 \noindent with the completeness relation given by
\begin{equation}
\sum_{\mbox{b}} \;{\cal U}^*_b(x - x_0) \, {\cal U}_b(x^\prime - x_0) + \int dk \;{\cal U}^*_k(x - x_0) \, {\cal U}_k(x^\prime - x_0)\,=\,\delta(x - x^\prime)
\end{equation}
and the subscript b stands for summation over bound states and k for
scattering states. For bound states, the eigenvectors are chosen to be real and for scattering states, we label them as ${\cal U}_k(x - x_0)$ and
are chosen such that ${\cal U}_k^\ast = {\cal U}_{-k}$, in which case the coordinate operators obey the hermiticity condition $q^*_k(t) = q_{-k}(t)$.

These eigenvectors are normalized as
\begin{equation}
\int {\rm d} x \, {\cal U}^\ast_p(x - \hat x_0)\, {\cal U}_q(x - \hat x_0) = \delta_{p,q}. 
\label{upuq}
\end{equation}

As a consequence of translational invariance, there is a mode with zero
eigenvalue given by\cite{c7}-\cite{lee}

\begin{equation} 
{\cal U}_0(x - x_0) = {1 \over{\sqrt M}} \left({{{\rm d} \phi_s}\over{{\rm d} x}}
\right) .
\label{zero}
\end{equation}

Depending on the particular form of the potential $U(\phi)$ there may
be other bound states (as is the case with the $\phi^4$ potential). There
is a continuum of scattering states with frequencies $\omega_k^2= k^2+\omega^2_o \; ; \; \omega^2_o = d^2U(\phi)/d^2\phi |_{\phi_{\infty}}$. These scattering states correspond  asymptotically to phase shifted plane waves in the cases under consideration because the relevant potentials are reflectionless\cite{c7,d2}. 
The continuum states are identified with meson states, whereas bound states
(other than the zero mode) are identified with excited states of the 
kink\cite{jackiwrevmod}.

The fluctuation along the functional direction corresponding to the zero frequency mode  represents an infinitesimal translation of the kink that costs no energy. Since this
mode has no restoring force, any arbitrarily large amplitude fluctuation along this direction is energetically allowed and therefore must be treated non-perturbatively. The variable $x_0$, i.e. the center of mass of the kink is elevated to the status of a quantum mechanical variable, and the fluctuations are orthogonal to the zero mode. 
This treatment is the basis of the collective coordinate
 method\cite{c7,lee,c1}-\cite{c22}.

In collective coordinates quantization  instead of the expansion (\ref{ex}) with (\ref{fluk}) we expand ${\phi}(x,t)$ as 
\begin{equation}
\phi(x,t) = \phi_s(x - \hat x_0(t)) + \sum_{n \neq 0}^\infty Q_n(t)\, {\cal U}_n(x - \hat x_0(t) )\;.
\label{phicc}
\end{equation}

This amounts to a change of basis in functional space, from the ``cartesian''
coordinates $\{ q_n \}$ to ``curvilinear'' coordinates $\{\hat{x}_0,Q_{n\neq 0} \}$\cite{c7,lee,c3,c6}. 

The next step is to express the Hamiltonian in terms of the new variables $\hat{x}_0(t)$ and $Q_n(t)$. For this we follow references\cite{c7,lee,c3,c6} and which we summarize below for the cases under consideration. 

\subsection{Kinetic and potential energies }

In the Schroedinger representation the kinetic energy can be expressed as a functional derivative as
\begin{equation}
T = -{1\over 2} \int {\rm d} x {\delta\over{\delta \phi}} {\delta 
\over{\delta \phi}}\; ,
\label{Tphi}
\end{equation}
where the functional derivative is written in the new coordinates using the chain-rule
\begin{equation}
{\delta \over{\delta \phi(x)}} = {\delta \hat x_0 \over{\delta \phi(x)}} {\delta 
\over{\delta \hat x_0}} + \sum_{m \neq 0} {\delta Q_m \over{\delta \phi(x)}} 
{\delta \over{\delta Q_m}}.
\label{ddelphi}
\end{equation}

Taking the functional variation of the field $\phi$, eq.(\ref{phicc}), we obtain
\begin{eqnarray}
\delta \phi(x) &=& {\delta \phi(x) \over{\delta \hat x_0}} \delta \hat x_0 +
\sum_{m \neq 0} {\delta \phi(x) \over{\delta Q_m}} \delta Q_m \nonumber \\
 &=& \left[ {\partial \phi_s(x - \hat x_0) \over{\partial 
\hat x_0}} +
\sum_{m\neq 0} Q_m {\partial {\cal U}_m(x - \hat x_0) \over{\partial \hat x_0}}\right] 
\delta \hat x_0 + \sum_{n \neq 0} {\cal U}_n(x - \hat x_0) \delta Q_n \; . 
\label{delphi}
\end{eqnarray}

Projecting both sides of the above equation on  ${\cal U}^\ast_0(x - \hat x_0)$ and then  ${\cal U}^\ast_p(x - \hat x_0)$ with $p\not=0$, using eqn.(\ref{zero}) and the orthonormalization condition eqn.(\ref{upuq}), we obtain: 
\begin{eqnarray}
{\delta \hat x_0 \over{\delta \phi(x)}} & =&  - {1\over{\sqrt M}} 
{1\over{\left[ 1 + (1/\sqrt M) \sum_{m \neq 0} Q_m S_m \right]}} 
{\cal U}^\ast_0(x - \hat x_0) \label{xx} \\
{\delta Q_p \over{\delta \phi(x)}}& = &  {\cal U}^\ast_p(x - \hat x_0) - 
{1\over{\sqrt M}} {{\sum_{n \neq 0} G_{pn}Q_n}\over{\left[ 1 
+ (1/\sqrt M) \sum_{m \neq 0} Q_m S_m \right]}} 
{\cal U}^\ast_0(x - \hat x_0), \label{cddphi}
\end{eqnarray}
where the matrix elements $G_{pm}$ are defined as
\begin{equation}
G_{pm} \,=\, \int {\rm d} x \,{\cal U}^\ast_p(x - \hat x_0)\,
\frac{\partial{\cal U}_m(x - \hat x_0)}{\partial x} \label{me1} 
\end{equation}
\begin{equation}
S_m \,\equiv\, G_{0m} =  \int {\rm d} x \,{\cal U}_0(x - \hat x_0) \,
\frac{\partial{\cal U}_m(x - \hat x_0)}{\partial x} \; .
\label{gmn}
\end{equation}

At this stage it is straightforward to follow the procedure detailed
in\cite{lee,c3,c6} to find the final form of the kinetic term in the Hamiltonian
in the Schroedinger representation of the coordinates 
$\hat{x}_0, Q_{m \neq 0}$:


\begin{eqnarray}
T &=& - {1\over 2} \left\{ {1\over{D}} {\delta \over{\delta \hat x_0}}
{\delta \over{\delta \hat x_0}} + {1\over{\sqrt D}} 
{\delta \over{\delta \hat x_0}} \sum_{p,m \neq 0}\left[ 
\frac{G_{pm}Q_m}{\sqrt{D}} {\delta \over{\delta Q_p}} + {\delta \over{\delta Q_p}} \frac{G_{pm}Q_m}{\sqrt{D}} \right] + \right. \nonumber \\
&& \quad\quad \left. {1\over{\sqrt D}} \sum_{p,q,m,n\neq 0} 
{\delta \over{\delta Q_p}}\left[\delta_{-p,q} \sqrt{D}\,+\,\frac{G_{pm}Q_m}{\sqrt{D}}G_{qn}Q_n
\right]{\delta \over{\delta Q_q}} \right\},
\label{Tqxf}
\end{eqnarray}

\noindent where $\sqrt D$ is the Jacobian associated with the change of coordinates\cite{c7,lee,c3,c6} and given by
\begin{equation}
\sqrt D \equiv \sqrt M \left[ 1+ \frac{1}{\sqrt M}\sum_{m \neq 0}Q_m S_m \right]. \label{jacobian}
\end{equation}

The total potential energy, including the elastic term, $V[\phi ]$ (see eqn.(\ref{hs})), is given by
\begin{equation}
V[\phi ] \equiv \int {\rm d} x \left[ {1\over 2} 
  \left({{\partial \phi}\over{\partial x}}\right)^2 + U(\phi) \right]\, .
\end{equation}

Using the expansion given by eqn.(\ref{phicc}) we find that it can be written in terms of the new coordinates as
\begin{equation}
V[\phi] = M + {1\over 2} \sum_{m \neq 0} Q_m Q_{-m} \omega_m^2 + {\cal O}(Q^3) +\cdots \,.\label{Vfin}
\end{equation}
By translational invariance the potential energy does not depend on the
collective coordinate. 
Identifying the canonical momenta conjugate to $\hat{x}_0, Q_n$ as 
\begin{equation}
\pi_0 \equiv P\;=\;-i {\delta \over{\delta \hat x_0}} \quad ; \quad
\pi_k = -i {\delta \over{\delta Q_{-k}}} \quad \hbox{for} \quad k \neq 0 \,.
\label{ppi}
\end{equation}

\noindent and using  the commutation relation of $\sqrt D \hbox{ and } 1/\sqrt D \,$ with $Q_n, \, \pi_n \hbox{ and } P$ given by
\begin{equation}
\left[\pi_n, \sqrt D\right] = -i S_n \quad \hbox{and} \quad ; \quad \left[\pi_n, {1\over{\sqrt D}}\right] = -i {S_n \over D}, 
\label{pisd}
\end{equation}
 we find the final form of the Hamiltonian:

\begin{eqnarray}
H &=& M + {1\over 2} \left\{ {P^2\over{D}} + {P\over{\sqrt D}} \sum_{p,m \neq 0}\left[ 
\frac{G_{pm}Q_m}{\sqrt{D}} \pi_{-p} + \pi_{-p} \frac{G_{pm}Q_m}{\sqrt{D}} \right] + \sum_{p \neq 0} \omega_p^2\,Q_p Q_{-p} \right. + \nonumber \\
&& \quad\quad \left. {1\over{\sqrt D}} \sum_{p,q,m,n\neq 0} 
\pi_{-p}\left[\delta_{-p,q} \sqrt{D}\,+\,\frac{G_{pm}Q_m}{\sqrt{D}}G_{qn}Q_n
\right]\pi_{-q} \right\} + {\cal O}(Q^3)+\cdots,
\end{eqnarray}
where $Q_p$ are now operators. The coordinates $Q_k$ associated with
the scattering states describe the meson  degrees of freedom with  frequency 
$\omega^2(k)=k^2+\omega^2_o\; ; \; \omega^2_o = d^2U(\phi)/d^2\phi|_{\phi_{\infty}}$. 
Since the Hamiltonian does not depend on $\hat{x}_0$ its canonical momentum $P$
is conserved, it is identified with the total momentum of the kink-meson
system\cite{c7,lee,c3}. The kink velocity, however, is not proportional to
$P$ and depends on the momentum of the meson field. 

Since our goal is to study the dynamics of the kink by obtaining the
equation of motion for the expectation value of the kink collective coordinate,
we introduce an external source term linearly coupled to $\hat{x}_0$. This
source term has a dual purpose, one is to allow to obtain the correlation function
of the collective coordinate as functional derivatives with respect to this
source, the other is to use this source as a Lagrange multiplier to turn the
evolution equation into an initial value problem. This second use will become
clear later when we study the solutions to the equations of motion. Therefore
we add the term $j(t) \hat{x}_0$ to the Hamiltonian.

\section{A DOMAIN WALL IN THE MESON HEAT BATH \label{secDMbath}}

Our goal is to study the dynamics of a domain wall in interaction with the mesons.
This is achieved by obtaining the real-time equations of motion of the collective 
coordinate $\hat{x}_0$ by  treating the mesons  as a ``bath'' and obtaining an influence functional\cite{d5}-\cite{d88} by ``tracing out'' the meson degrees of freedom and the excited states of the kink. 
 We assume that the total density matrix for the
kink-meson system decouples at the initial time $t_i$, i.e.
\begin{equation}
\rho(t_i)\,=\,\rho_s(t_i)\,\otimes\,\rho_R(t_i),
\end{equation}
where $\rho_s(t_i)$ is the density matrix of the system which is taken to be 
that of a free particle associated with the collective coordinate of the
kink, i.e. $\rho_s(t_i)= |x_0><x_0|$ and $\rho_R(t_i)$ is the density matrix of the meson bath and describes  mesons in thermal equilibrium at a temperature $T$.

Since the kinks can never be separated from the meson fluctuations, this factorization must be understood to hold in the limit in which the
initial time $t_i \rightarrow -\infty$ with an adiabatic switching of the
kink-meson interaction.

The time evolution is completely contained in the time dependent density matrix
\begin{equation}
\rho(t)= U(t,t_i)\rho(t_i)U^{-1}(t,t_i)\label{rhotime}
\end{equation}
with $U(t,t_i)$ the time evolution operator. Real time non-equilibrium expectation
values and correlation functions can be obtained via functional derivatives
with respect to sources of the generating functional\cite{b1}-\cite{b6}
\begin{equation}
Z[j^+,j^-] = Tr U(\infty,-\infty;j^+)\rho_i U^{-1}(\infty,-\infty;j^-)/Tr\rho(t_i),
\label{genefunc}
\end{equation}
where $j^{\pm}$ are sources coupled to the meson field and the collective 
coordinate. This generating functional  is readily obtained using the Schwinger-Keldysh method which involves a path integral in a complex contour in time\cite{b1}-\cite{b7}: a branch corresponding to the
time evolution forward, a backward branch corresponding to the inverse time
evolution operator and a branch along the imaginary time axis
from $t_i$ to $t_i-i\beta$ to represent the
initial thermal density matrix. 
We will obtain the equation of motion for the kink collective coordinate
in an expansion of the ``adiabatic'' parameter $\omega_0/M$which 
is also the weak coupling limit of the scalar field theories under consideration
\cite{c7}.
As it will be shown explicitly below in the particular cases studied, the 
matrix elements given by eqns.(\ref{me1},\ref{gmn}) will provide the necessary powers of the meson mass $\omega_0$. The lowest order in $\omega_0/M$ is formally 
obtained by  keeping only the $1/M$ terms in the Hamiltonian and 
neglecting the non-linear ${\cal O}(Q^3)$ terms. Under these approximations, $1/D \simeq 1/M$ and the Hamiltonian has the following form
\begin{eqnarray}
H =  M \,+ \,{1\over{2 M}} \left( P + \sum_{m,n \neq 0} D_{mn}\pi_m\, Q_n
\right)^2 + \frac{1}{2}\sum_{m \neq 0}\left[ \pi_m \pi_{-m} + \omega_m^2 Q_m 
Q_{-m} \right] + {j}(t) \hat{x}_0,
\label{Happrof}
\end{eqnarray}
where we define 
\begin{equation}
D_{mn} = G_{-mn}.
\label{Dmn}
\end{equation}

At this point it proves convenient to write the coordinates and momenta 
of the mesons in terms of creation and annihilation operators obeying the
standard Bose commutation relations,
\begin{equation}
Q_k = {1\over{\sqrt{2\omega_k}}} \left[ a_k + a_{-k}^\dagger \right] 
\; \; ; \; \; 
\pi_k = -i \sqrt{{\omega_k \over 2}} \left[ a_k - a_{-k}^\dagger\right]\,.
\label{QP}
\end{equation}
 
The Hamiltonian can be expressed in terms of $a$ and $a^\dagger$ as
\begin{equation}
H = {1\over{2 M}} \left( P + F[a^\dagger, a]\right)^2 + 
\sum_{k \neq 0} \omega_k \left(a_k^\dagger \,a_k + 1/2\right) +
{j}(t)\hat{x}_0+ M, 
\label{Haa}
\end{equation}
where 
\begin{equation}
F[a^\dagger, a] = \sum_{p,k\not=0} \left[T^{(S)}_{pk}\,\left(a_p\,a_k \,-\,a^\dagger_{-p}\,
a^\dagger_{-k}\right)\,+\,T^{(A)}_{pk}\,\left(a^\dagger_{-p}\,a_k \,-\,a^\dagger_{-k}\,
a_{p}\right) \right].
\label{Faa}
\end{equation}

We have made use of the symmetries of the operators and defined the symmetric $T^{(S)}_{pk}$ and antisymmetric $T^{(A)}_{pk}$ matrices that
provide the interaction vertices as
\begin{equation}
T_{kp}^{(S)} = {1\over{4 i}} \left[\sqrt{{\omega_k
\over{\omega_p}}} - \sqrt{{\omega_p\over{\omega_k}}}\, 
\right] D_{kp} \quad ; \quad
T_{kp}^{(A)} = {1\over{4 i}} \left[\sqrt{{\omega_k
\over{\omega_p}}} + \sqrt{{\omega_p\over{\omega_k}}}\, 
\right] D_{kp}. \label{scm}
\end{equation}

 To use the path integral formulation we need the Lagrangian, which to the order that we are working (${\cal O}(\omega_0/M)$) and properly accounting for
normal ordering, is given by
\begin{equation}
{\cal L} [\dot{\hat x_0},a,a^\dagger ] = {M\over 2} \dot{\hat x_0}^2 - \dot{\hat
x_0} F[a^\dagger, a] - \sum_{k \neq 0} \omega_k (a_k^\dagger a_k + 1/2) -
{j}(t)\hat{x}_0- M.
\label{lag}
\end{equation}

The interaction of the collective coordinate and the mesons 
is now clear. Only time derivatives of the collective coordinate couple, a 
consequence of the Goldstone character of the collective coordinate. There
are two processes described by the interaction: i) creation and destruction of
two mesons and ii) scattering of mesons. Whereas the first type can contribute
with the mesons in their ground state, the second can only contribute if 
the meson states are occupied. The two processes are depicted in fig.(\ref{bubbles}). As it will become clear below, the second type of processes will lead to Landau damping.

Since we have preferred to work in terms of the creation and annihilation
operators it is convenient to write the path integral for the non-equilibrium generating functional in the coherent state representation\cite{neto,negele}.

 Following the steps outlined in reference\cite{neto,negele} we find the generating functional
of non-equilibrium Green's functions in the coherent state representation to be given by

\begin{equation}
{\cal Z}[j^+,j^-] =  \int {\cal D}x^+ \int {\cal D}x^- \int {\cal D}^2 \gamma^+   \int {\cal D}^2 \gamma^-  exp\left\{i\int dt \left({\cal L}[\dot{x}^+,\gamma^{*+},\gamma^+,j^+]\,-\,{\cal L}[\dot{x}^-,\gamma^-,
\gamma^{*-},j^-]\right)\right\}
\label{trgnr}
\end{equation}
with the Lagrangian density defined on each branch given by
\begin{eqnarray} 
{\cal L}[\dot{x}^{\pm},\gamma^{\pm},\gamma^{*\pm},j^{\pm}] & = &  {M \over 2} 
\left( \dot{x}^{\pm}\right)^2  + \sum_{k\neq 0}\left[i\gamma^{*\pm}_k 
\frac{d \gamma^{\pm}_k}{dt} - \omega_k \gamma_k^{*\pm} \gamma_k^{\pm}+\gamma_k^{\pm} j^{*\pm}_k+ \gamma_k^{*\pm} j^{\pm}_k\right] - \nonumber \\
& &  \dot{x}^{\pm} F[\gamma^{* \pm}, \gamma^{\pm}] -{j}(t)x^{\pm}
\label{lagnoneq}
\end{eqnarray}
and with proper boundary conditions on the fields that reflect the factorized 
initial condition with the mesons in thermal equilibrium.  The signs $\pm$ in the above expressions correspond to the fields and sources on
the forward ($+$) and backward $(-)$ branches. The contribution from the branch along the
imaginary time is cancelled by the normalization factor. This is
the non-equilibrium generalization of the coherent state path integrals (see details in appendix \ref{A}).
Non-equilibrium Green's functions are now obtained as functional derivatives
with respect to the sources $j^{\pm}$. There are 4 types of free meson propagators\cite{b1}-\cite{b6}: 
\begin{eqnarray}
< a_k^{\dagger +}(t) a_p^+(t^\prime)> &=& \delta_{k,p} 
e^{-i\omega_k(t^\prime - t)} [\theta(t^\prime - t) + n_k ] \nonumber \\ 
< a_k^+(\tau) a_p^{\dagger +}(\tau^\prime)> &=& \delta_{k,p} 
e^{i\omega_k(t^\prime - t)} [\theta(t - t^\prime) 
+ n_k ] \nonumber \\ 
< a_k^{\dagger (\pm)}(t) a_p^{\dagger (\pm)}(t^\prime)> &=& 0 
\nonumber \\ 
< a_k^{(\pm)}(t) a^{(\pm)}_p(t^\prime)> &=& 0 \\ 
< a_k^{\dagger +}(\tau) a_p^-(\tau^\prime)> &=& \delta_{k,p} 
e^{-i\omega_k(t^\prime - t)} [1 + n_k ] \nonumber \\
< a_k^+(\tau) a_p^{\dagger -}(\tau^\prime)> &=& \delta_{k,p} 
e^{i\omega_k(t^\prime - t)} n_k,  \nonumber 
\label{propaa}
\end{eqnarray}
where $n_k$ is Bose Einstein distribution for mesons of quantum number $k$ and
$<\cdots>$ refer to averages in the initial density matrix. The
$++$ ($--$) propagators correspond to the time-ordered (anti-time-ordered), whereas the $\pm \mp$ are the Wightman functions.

An important point to notice is that 
\begin{equation}
< F[a^\dagger,a]> \,=\,0 \label{faad}
\end{equation}
in the non-interacting case, since it is proportional to $\sum_k D_{k,-k}=0$.

\subsection{The equation of motion for the collective coordinate}

The equation of motion of the expectation value of the collective coordinate for the kink $<\hat{x}_0>=q(t)$,  can be derived by expanding $x^{\pm}(t)\,=\,
q(t)\,+\,\xi^{\pm}(t)$ and requiring $<\xi^{\pm}(t)>\,=\,0$ to all orders in  perturbation theory\cite{data}. 
Imposing the condition $<\xi^+(\tau^\prime)>\,=\,0$, treating the interaction
term  up to second order in perturbation theory  and using eqn.(\ref{faad}), we obtain the following linearized equation of motion
\begin{equation}
\int_{-\infty}^{\infty} dt^\prime <\xi^+(t)\dot{\xi}^+(t^\prime)>\left[
 \left\{ M\dot{q}(t^\prime) +
\int_{-\infty}^t dt^{\prime\prime} \Gamma_m(t^\prime-t^{\prime\prime}) \dot{q}(t^{\prime\prime}) \right\} + <\xi^+(t)\xi^+(t^\prime)> {j}(t^\prime)\right]\,=\,0 ,
\label{eqnms}
\end{equation}

\noindent where the retarded kernel is given by
\begin{eqnarray}
-i\,\Gamma_m(t-t^{\prime})\theta(t-t^{\prime}) & = & <F[a^{\dagger +}(t),a^+(t)]\;F[a^{\dagger +}(t^\prime),a^+(t^\prime)]> \nonumber \\
& - & <F[a^{\dagger +}(t),a^+(t)]\;F[a^{\dagger -}(t^\prime).
a^-(t^\prime)]> \label{GA}
\end{eqnarray}

Since we restrict ourselves to non-relativistic kinks we consider $\dot{q}\ll1$. The non-equilibrium Feynman diagrams that contribute to one loop order (second order) are
shown in fig.(\ref{bubbles}). 

Alternatively this equation of motion may be obtained by computing the
influence functional\cite{d5}-\cite{d88} in second order perturbation theory. The resulting
influence functional is quadratic in the collective coordinate,  performing the shift $x^{\pm}(t) = q(t)+\xi^{\pm}(t)$ the above equation of motion is obtained
by requesting that the linear terms in $\xi^{\pm}$ vanish, (there are
two linear terms, both give the same equation of motion).

The kernel $\Gamma_m(t-t^\prime)$ is found by using  eqn.(\ref{GA}) and eqn.(\ref{propaa})  and it is given by 
\begin{eqnarray}
\Gamma_m(t - t^\prime) &=& -4\,  \sum_{p,k\not=0} \left\{ T_{pk}^{(S)}\, 
T_{-p-k}^{(S)} (1 \,+ 2 n_p) \,\sin \left[ (\omega_p + \omega_k)
(t - t^\prime)\right] \right. \nonumber \\
&& \left. -  \quad 2 T_{pk}^{(A)} \, T_{-p-k}^{(A)} n_p 
\sin\left[(\omega_p - \omega_k)(t - t^\prime)\right] \right\}.
\label{GTT}
\end{eqnarray}

Performing the integral over $t^\prime$ in eqn.(\ref{eqnms}) by parts, we obtain the final form of the equation of motion 
\begin{equation}
M \ddot{q} (t) + \int_{-\infty}^t {\rm d} t^\prime \,
\Sigma_m(t-t^\prime)\dot{q}(t^\prime) = {j}(t),
\end{equation}
where the non-local kernel is given by
\begin{equation}
\Sigma_m(t-t^\prime) = {\partial \Gamma_m(t-t^\prime) 
\over{\partial t}}= -{\partial \Gamma_m(t-t^\prime) 
\over{\partial t^{\prime}}}\,.
\label{Sigm}
\end{equation} 

Using eqn.(\ref{GTT}) we find the final expression for the kernel $\Sigma_m$:
\begin{eqnarray}
\Sigma_m(t-t^\prime) &=&  -4 \,  \sum_{p,k\not=0} \left\{ T_{pk}^{(S)} \,
T_{-p-k}^{(S)} (1 +  2n_p ) (\omega_p + \omega_k) 
\cos \left[ (\omega_p + \omega_k)(t - t^\prime)\right]  
\right. \nonumber \\
&&\left. - 2\;T_{pk}^{(A)}\,T_{-p-k}^{(A)} n_p \,
(\omega_p - \omega_k) \cos\left[(\omega_p - \omega_k)
(t - t^\prime)\right] \right\}.
\label{sigmf}
\end{eqnarray}
We will see in the next sections that the two kernels $\Sigma_m \; ; \; \Gamma_m$
have very special significance: whereas $\Sigma_m$ is identified with
the real-time retarded self-energy of the collective coordinate, $\Gamma_m$ will provide the
coefficient of {\em dynamical friction} in the Markovian approximation.

It is more convenient to express the equation of motion of the kink in terms of the velocity
\begin{equation}
M \dot{V} (t) + \int_{-\infty}^t {\rm d} t^\prime \,
\Sigma_m(t-t^\prime)\,V(t^\prime) = {j}(t)
\label{eqmotm}
\end{equation}
with $\Sigma_m$ given by eqn.(\ref{sigmf}).

The relation (\ref{sigmf}), ensures to this order in the perturbative
expansion,  that
with an adiabatic switching on convergence factor introduced to regularize
the lower limit of the integral and to provide an initial factorization
of the density matrix as $t_i \rightarrow -\infty$ the
total integral of the retarded self-energy kernel vanishes, i.e,
\begin{equation}
\int_{-\infty}^t \Sigma_m(t-t')dt' = 0. \label{zeroint}
\end{equation}
This result is consistent with that found in Refs. \cite{khleb,arnold}.

Therefore for ${j}=0$, any  constant velocity of the kink is a solution of the equation of motion (\ref{eqmotm}). This result is physically clear: when $j=0$,
the total Hamiltonian commutes with $P$, the canonical momentum conjugate to
$\hat{x}_0$ because of translational invariance, i.e. the total momentum of
the system is conserved. One can then go to a frame in which $P=0$ and since
the meson bath is in equilibrium this must result in that the domain wall must have
a constant velocity, therefore for $j=0$ there must be a constant velocity solution to the equations of motion of the collective coordinate resulting in (\ref{zeroint}).

\subsection{General properties of the solution \label{gensol}}

 Since in the absence of an external driving term we have
found that the domain wall moves with constant velocity, we can now use the external source
term to cast the evolution as an initial value problem. For this consider the situation in which at time $t=0$ a force is applied, therefore changing the velocity of the domain wall.
 Assuming that for $t<0$ the kink traveled with a constant velocity $v_0$,
 after switching on the external  force the domain wall will accelerate, but it will also transfer energy and
excite the meson degrees of freedom and this will lead to dissipative
processes. Therefore writing $V(t)=v_0+ v(t)$ with ${j}(t<0)=0 \;; \; {j}(t>0) \neq 0$ and using the property (\ref{zeroint}) the
equation of motion for the velocity change becomes
\begin{equation}
 M \dot{v} (t) + \int_{0}^t {\rm d} t^\prime \,
\Sigma_m(t-t^\prime)\,v(t^\prime) = {j}(t).\label{eqnofmotlap}
\end{equation}

The solution of this equation is found by Laplace transform,
in terms of $\tilde{v}(s)\; ; \;  \tilde {\Sigma}_m(s)\; ; \; \tilde{J}(s)$,
the Laplace transforms of the velocity, self-energy kernel and current
respectively, in terms of the Laplace variable $s$. We find that the
solution is given by
\begin{equation}
\tilde{v}(s) = \frac{v_0+(\tilde{J}(s)/M)}{s+\frac{1}{M}\tilde{\Sigma}_m(s)},
\label{laplasol}
\end{equation}
and consistently with the linearized equation of motion and the non-relativistic dynamics
$v(t); v_0 \ll 1$.
The quantity 
\begin{equation}
G(s) = \frac{1}{s+\frac{1}{M}\tilde{\Sigma}_m(s)}
\label{propag}
\end{equation}
is the Laplace transform of the propagator of the velocity of the collective coordinate.
The real time evolution is found by the inverse Laplace transform
\begin{equation}
v(t) = \frac{1}{2\pi i} \int_C e^{st} \tilde{v}(s) ds
\end{equation}
where $C$ refers to the Bromwich contour running along the imaginary
axis to the right of all the singularities of $\tilde{v}(s)$ in the
complex $s$ plane. Therefore we need to understand the analytic 
structure  of $G(s)$ in eqn. (\ref{laplasol}) to obtain the real time dynamics. 
The Laplace transform of the self-energy kernel is conveniently
written as a dispersion relation in the form

\begin{eqnarray}
\tilde{\Sigma}_m(s) & = & s \tilde{\Gamma}_m(s) \label{lapfric} \\
\tilde{\Gamma}_m(s) & = & \int \frac{\rho(p_o)}{s^2+p^2_o} dp_o
\label{dispersionrel} \\
\rho(p_o) & = & 
-4\,  \sum_{p,k\not=0} \left\{ T_{pk}^{(S)}\, 
T_{-p-k}^{(S)} (1 \,+ 2 n_p) \,\delta(p_o-\omega_p - \omega_k)
 \right. \nonumber \\
&  & \left. -  \quad  T_{pk}^{(A)} \, T_{-p-k}^{(A)} (n_p-n_k) 
\delta(p_o-\omega_p + \omega_k)
\right\} \label{spectralrho},
\end{eqnarray}

 where $\tilde{\Gamma}_m(s)$ is the Laplace transform of the kernel
$\tilde{\Gamma}_m$ given above. 

This dispersive form for the Laplace transform of the kernel reveals that $\tilde{\Gamma}_m(s)$
has a discontinuity in the complex s-plane along the imaginary axis, since
\begin{equation}
\tilde{\Gamma}_{mI}(s=i\omega + 0^{\pm}) = \mp \frac{\pi sign(\omega)}{2\,|\omega|} \left[\rho(|\omega|)-
\rho(-|\omega|)\right]  \label{disconti}
\end{equation} 

The imaginary part changes sign with $\omega$ as a result of  the retarded nature of the kernel. 
Therefore the propagator $G(s)$ has cuts along the imaginary axis in the complex s-plane.
The two different contributions to the spectral density (\ref{spectralrho}) yield to two
different cut structures. For $\omega >0$, the first term, proportional to 
$\delta(\omega-\omega_p - \omega_k)$ gives a two-meson cut beginning at $2\omega_0$ corresponding
to the process of spontaneous and induced two-meson creation and annihilation. The second 
contribution corresponding
to $\delta(\omega-\omega_p + \omega_k)$ gives a cut, which we identify as the Landau damping cut\cite{weldon,data}, pinching the origin and originates
in the process of scattering of mesons present in the medium off the domain wall. 
As it will be seen in detail for the examples in the next section the discontinuity vanishes linearly as $\omega \rightarrow 0$ allowing an analytic
continuation into the second Riemann sheet and to isolate the pole. This linear vanishing of the
self energy is consistent with the case studied by\cite{khleb,arnold}.
This 
process is present only for finite temperature as there must be mesons present for this
process to exist. This contribution is identified as  Landau damping  from
the in medium mesons and will be seen to provide the leading contribution to the long time relaxation.

The presence of a static friction coefficient will be revealed by a
pole in $G(s)$ with a negative real part, since this will translate into 
an exponential relaxation of the velocity.

In the absence of
interactions $G(s)$ has a simple pole at $s=0$. Since we obtained the
expression for the kernels in perturbation theory the position of a
pole must be found in a consistent perturbative expansion by writing
$s_p= (1/M) s_1 + \cdots $, we find
\begin{equation}
s_p = -\frac{1}{M}\,\tilde{\Sigma}_m(s=0) \equiv 0. \label{staticfric}
\end{equation} 
 Therefore the coefficient of {\em static} friction vanishes.
This is a consequence of the vanishing of the integral (\ref{zeroint}). 
Therefore up to this order in perturbation theory the position of the
pole in the s-variable remains at $s=0$ resulting in that the {\em static} friction
coefficient vanishes.

 In summary, the analytic structure of $G(s)$ in the complex 
s-plane corresponds to a pole at $s=0$ with residue
\begin{equation}
Z_s = \frac{1}{1+\frac{1}{M}\tilde{\Gamma}_m(0)} \label{wavefunc}
\end{equation} 
and cuts along the imaginary axis beginning at $\pm 2i \omega_o \; ; \pm i\epsilon$ 
with $\epsilon \rightarrow 0$ to clarify that the beginning of this cut pinches the pole at the origin but the continuum contribution to the spectral density
(discontinuity) vanishes at the position of the pole at $s=0$.

The residue $Z_s$ has a very clear interpretation, it is the  ``wave function renormalization'' and its effect can be understood in
two alternative manners.

Consider the case in which $\tilde{J}=0$ in eqn.(\ref{laplasol}). 
Performing the inverse Laplace transform and invoking the Riemann-Lebesgue lemma,
the long time behavior will be completely dominated by the pole at $s=0$. Therefore, if the velocity of the kink has been changed at $t=0$ by some
external source, this disturbance will relax in time to an asymptotic value given by
\begin{equation}
v_{\infty} = Z_s v_0. \label{asinto}
\end{equation}

Alternatively, consider the case of $v_0=0$ but with an external source term switched on
at $t=0$ and constant in time thereafter. Again the inverse Laplace transform at long time will be dominated by the pole,  and we find that the kink moves
with constant acceleration given by
\begin{equation}
\dot{v}  = \frac{\tilde{j}}{M_{eff}} \quad \mbox{ with } \quad 
M_{eff}  =  \frac{M}{Z_s}. 
\label{acceleration} 
\end{equation}
Thus the wave function renormalization can also be understood as a renormalization 
of the kink mass. The ratio of the asymptotic acceleration to the
initial acceleration is given by $Z_s$. As the kink moves, the
interaction with the meson  bath ``dress'' it changing its 
effective mass, which will be seen in specific models  to be {\em larger} than the bare mass.

Furthermore we can now derive the following important sum rule. 
Consider the case $j \equiv 0$. Isolating the contribution from the
pole and the cuts (by replacing 
$\delta(p_o-\omega_p + \omega_k)\rightarrow \delta(p_o-\omega_p + \omega_k-\epsilon)$ and taking the limit $\epsilon \rightarrow 0$ at
the end of the computation) separating the pole and continuum contributions, and
deforming the contour of integration for the inverse Laplace transform as
shown in fig. (\ref{cut}) we find the time evolution (for $j=0$) to
be given by
\begin{equation}
\frac{v(t)}{v_0} =  Z_s+ \frac{2}{\pi M}\int_{\epsilon}^{\infty} \frac{d\omega}{\omega}
\frac{\left[-\tilde{\Gamma}_{Im}(\omega)\right]
\cos(\omega t)}{ \left[
1+\frac{\tilde{\Gamma}_{Rm}(\omega)}{M}\right]^2+ 
\left[
\frac{\tilde{\Gamma}_{Im}(\omega)}{M}\right]^2} \label{voftime}
\end{equation}
Evaluating at $t=0$ we obtain the sum rule
\begin{equation}
Z_s+ \frac{2}{\pi M}\int_{\epsilon}^{\infty}\frac{d\omega}{\omega}
\frac{\left[-\tilde{\Gamma}_{Im}(\omega)\right]}{\left[
1+\frac{\tilde{\Gamma}_{Rm}(\omega)}{M}\right]^2+ 
\left[
\frac{\tilde{\Gamma}_{Im}(\omega)}{M}\right]^2}  =1
\label{sumrule}
\end{equation}
Since the spectral density $\rho(\omega)$ is positive (semi)definite
as it will be explicitly shown below for specific models, the sum rule
above determines that 
\begin{equation}
Z_s <1 \Longrightarrow \frac{v_{\infty}}{v_0} < 1 \label{asimptorel}
\end{equation}

Although a sum rule similar to eqn.(\ref{sumrule}) is obtained in
quantum field theory from the canonical commutation relations, its 
validity for the collective coordinate associated with the domain wall
is far from obvious since the kinematic and canonical momentum conjugate to the collective coordinate are different. 

The continuum contribution in eqn.(\ref{voftime}) is dominated at
long times by the small $\omega$ region. Therefore for $T\neq 0$
 the asymptotic long time relaxation of the velocity is completely determined by the Landau damping cut which has support at small $\omega$, whereas the
two-meson cut vanishes below the threshold at $2\omega_0$.

A further understanding of the dynamics will necessarily require knowledge of the matrix elements to establish the details of the kernels. This will be studied in particular models in the next section.

\subsection{Semiclassical Langevin equation}

The classical Langevin equation is an adequate phenomenological description of Brownian motion obtained by considering the dynamics of one (or few) degrees of freedom that interact with a bath in equilibrium. 

 It contains a term proportional to the velocity of the particle which incorporates friction and dissipation and a stochastic term which reflects the random interaction of the heat bath with the particle. These two terms are related by the classical fluctuation-dissipation relation which is
derived in appendix \ref{C} (see eqn. (\ref{flucdiss}). 

At the quantum mechanical level it is also possible to obtain  a ``reduced'' or coarse grained description of the dynamics of one (or few) degrees of freedom
in interaction with a bath. The coarse graining procedure has a very precise
meaning: the full time dependent density matrix is traced over the bath degrees of freedom yielding an effective or ``reduced'' density matrix for the degrees of freedom whose dynamics is studied.

 Such a  description of non-equilibrium dynamics of a quantum mechanical particle coupled to a dissipative environment by a Langevin equation was presented by Caldeira and Leggett\cite{caldeira} and by Schmid \cite{d6}. Their technique is based on the influence-functional method of Feynman and Vernon \cite{d5} that naturally leads to a semiclassical  Langevin equation.

In this section we follow the procedure of\cite{caldeira}-\cite{d88} 
generalized to our case to derive the Langevin equation for kinks in a heat bath to lowest order in the adiabatic (weak) coupling. 

The main step is to perform the path integrals over the meson degrees of
freedom, thus obtaining an effective functional for the collective coordinate
of the kink. Unlike the most usually studied cases of a particle linearly coupled to an harmonic reservoir\cite{caldeira}-\cite{d88} we have here a bilinear coupling to the mesons. Therefore the influence functional cannot be obtained exactly,
but it can be obtained in a consistent perturbative expansion. For this we
treat  the interaction term ${\cal L}_I[\dot{x}^{\pm},\gamma^{\pm},\gamma^{*\pm}]$ in perturbation theory up to second order in the vertex proportional to $\dot{x}^{\pm}$ (which is equivalent to lowest order
in the adiabatic coupling $m/M$). Integrating over the meson variables and using $<F[a^\dagger,a]>=0$, we obtain
\begin{equation}
{\cal Z}[j^+,j^-=0] \,=\, \int {\cal D}x^+\,{\cal D}x^-\, \;e^{i\int_{-\infty}^{\infty} dt^\prime \left({\cal L}_0[\dot{x}^+]\,-\,{\cal L}_0[\dot{x}^-]\right)} \;{\cal F}[\dot{x}^+,\dot{x}^-],
\label{infl}
\end{equation}
where
\begin{equation}
{\cal L}_0[\dot{x}^\pm] \,=\, \frac{1}{2} M \left(\dot{x}^\pm\right)^2 - {j}
x^{\pm}
\label{L0}
\end{equation}
and ${\cal F}[\dot{x}^+,\dot{x}^-]$ is the influence 
functional\cite{d5}-\cite{d88}. To
lowest adiabatic order we find
\begin{eqnarray}
{\cal F}[\dot{x}^+,\dot{x}^-] & = & exp\left\{-\, \frac{1}{2} \int dt\, dt^\prime \right. \biggl[
\dot{x}^+(t) \, G^{++}(t,t^\prime) \, \dot{x}^+(t^\prime) \,+\, \dot{x}^-(t) \, G^{--}(t,t^\prime) \, \dot{x}^-(t^\prime) \nonumber \\
&  & \hspace{1in} + \; \left.  \dot{x}^+(t) \,G^{+-}(t,t^\prime) \,\dot{x}^-(t^\prime) \,+\,
\dot{x}^-(t) \, G^{-+}(t,t^\prime) \, \dot{x}^+(t^\prime) \biggr] \right\}
\end{eqnarray}
in terms of the real-time meson correlation functions  (see appendix (\ref{B}))
\begin{eqnarray}
G^{++}(t,t^\prime) & = &  <F[a^{\dagger +}(t),a^+(t)]\;F[a^{\dagger +}(t^\prime),a^+(t^\prime)]> \nonumber \\
G^{--}(t,t^\prime) & = &  <F[a^{\dagger -}(t),a^-(t)]\;F[a^{\dagger -}(t^\prime),a^-(t^\prime)]>\nonumber \\
G^{+-}(t,t^\prime) & = &  -\,<F[a^{\dagger +}(t),a^+(t)]\;F[a^{\dagger -}(t^\prime),a^-(t^\prime)]>\nonumber \\
G^{-+}(t,t^\prime) & = &  -\,<F[a^{\dagger -}(t),a^-(t)]\;F[a^{\dagger +}(t^\prime),a^+(t^\prime)]>.
\label{greens}
\end{eqnarray}

At this stage it is convenient to introduce the center of mass and relative coordinates, $x$ and $R$ respectively, which are defined as
\begin{equation}
x(t)\,=\, \frac{1}{2} \left( x^+(t)\,+\,x^-(t) \right) \hspace{.3in},\hspace{.3in}
R(t)\,=\,x^+(t)\,-\,x^-(t).
\label{wigner}
\end{equation}
These are recognized as the coordinates used in the Wigner transform of 
the density matrix \cite{caldeira}-\cite{d88} in terms of which the partition function becomes
\begin{equation}
{\cal Z}[0] \,=\, \int {\cal D}x\,{\cal D}R\, \;e^{iS[x,R]}
\label{p2}
\end{equation}
with the non-equilibrium effective action given by 
\begin{equation}
S[x,R] \,=\, \int dt \, R(t) \, \left[ -M \ddot{x}(t) \,-\, \frac{i}{2}\int dt^\prime\,\biggl(
K_1(t-t^\prime)\, \dot{x}(t^\prime) \,-\,K(t-t^\prime)\,R(t^\prime) \biggr) \right] 
\label{action}
\end{equation}
in terms of the kernels $K_1(t-t^\prime)$ and $K(t-t^\prime)$ which are given by (see appendix (\ref{B}) )
\begin{eqnarray}
K_1(t-t^\prime) & = & 8\,i\, \theta(t-t^\prime) \sum_{p,k\not=0} \biggl\{ T_{pk}^{(S)} \,
T_{-p-k}^{(S)} (1 +  n_p + n_k ) (\omega_p + \omega_k) 
\cos \left[ (\omega_p + \omega_k)(t - t^\prime)\right]    \nonumber \\
&& - \;T_{pk}^{(A)}\,T_{-p-k}^{(A)}( n_p - n_k )\,
(\omega_p - \omega_k) \cos\left[(\omega_p - \omega_k)
(t - t^\prime)\right] \biggr\} \nonumber \\
& = & -2\,i\,\Sigma_m(t-t^\prime)
\label{ker1}
\end{eqnarray}
and
\begin{eqnarray} 
K(t-t^\prime) & = & -2\, \sum_{p,k\not=0} \biggl\{ T_{pk}^{(S)} \,
T_{-p-k}^{(S)} (1 +  n_p + n_k + n_p\,n_k) (\omega_p + \omega_k)^2 
\cos \left[ (\omega_p + \omega_k)(t - t^\prime)\right]  
 \nonumber \\
&& + \,2\, \;T_{pk}^{(A)}\,T_{-p-k}^{(A)} n_k( 1 + n_p )\,
(\omega_p - \omega_k)^2 \cos\left[(\omega_p - \omega_k)
(t - t^\prime)\right] \biggr\}. \label{ker}
\end{eqnarray}

At this stage it proves convenient to introduce the identity 
\begin{eqnarray}
e^{-\frac{1}{2}\int dt \, dt^\prime R(t) \; K(t-t^\prime)\, R(t^\prime)} & = & C(t) \int {\cal D} \xi \;e^{-\frac{1}{2} \int dt \, dt^\prime  \xi(t)K^{-1}(t-t^\prime)\xi(t^\prime) \,+\, i\int dt\,\xi(t) R(t) }
\end{eqnarray}
with C(t) being an inessential normalization factor, to cast the 
 non-equilibrium effective action of the collective coordinate
in terms of a stochastic noise variable with a definite probability
 distribution\cite{d6}-\cite{d88}.

\begin{eqnarray}
{\cal Z}[0] &=& \int {\cal D}x\,{\cal D}R\,{\cal D} \xi \,P[\xi]\; exp\biggl\{i \int dt R(t) \biggl[ -M \ddot{x}(t) \,-\, \frac{i}{2}\int dt^\prime\,
K_1(t-t^\prime)\, \dot{x}(t^\prime) \,+\, \xi(t) \biggr] \biggr\}, \nonumber \\
\label{ZLang}
\end{eqnarray}
where the probability distribution of the stochastic noise, $P[\xi]$, is given by
\begin{equation}
P[\xi]\,=\,\int {\cal D} \xi \;exp\left\{ -\frac{1}{2} \int dt \, dt^\prime  \xi(t)K^{-1}(t-t^\prime)\xi(t^\prime) \right\}.
\end{equation}
 In this approximation we find that the noise is Gaussian, additive and with  correlation function  given by
\begin{equation}
< \xi(t) \xi(t^\prime)>\,=\, K(t-t^\prime).
\end{equation}

The semiclassical Langevin equation is obtained by extremizing  the effective action  in eqn.(\ref{ZLang})with respect to $R(t)$\cite{caldeira}-\cite{d88}
\begin{equation}
M \ddot{x}(t) \,+\, \int_{-\infty}^t dt^\prime\,
\Sigma_m(t-t^\prime)\, \dot{x}(t^\prime)-{j}(t) \,=\, \xi(t).
\label{Lang}
\end{equation}

Two features of the semiclassical Langevin equation  deserve comment. The first is that the kernel $K_1(t-t^\prime)$, as can be seen from eqn.(\ref{ker1}), is non-Markovian. The second is that the noise correlation function $K(t-t^\prime)$ is colored, i.e. it is not a delta function $\delta(t-t^\prime)$. The relationship between the kernels $K_1(t-t^\prime)$ and $K(t-t^\prime)$ established in appendix (\ref{C}) constitutes a generalized quantum fluctuation dissipation 
relation\cite{caldeira}-\cite{d88} (see eqn.(\ref{flucdiss}). 
Finally we recognize that taking the average of (\ref{Lang}) with the
noise probability distribution $P[\xi]$ yields the equation of motion for
the expectation value of the collective coordinate (eqn.(\ref{eqmotm})).

A classical description is expected to emerge when the occupation distribution for the
mesons can be approximated by their classical counterparts\cite{caldeira}, i.e. when $n_k \approx T/\omega_k$.

{\em If} the kernels $\Sigma_m$ and $K$ admit a Markovian limit then a diffusion 
coefficient could be extracted by computing the long time limit of the 
correlation function $<<(x(t)-x(0))^2>>/t$ where $<<\cdots>>$ stand for average over the noise distribution function. However when the kernels do not become Markovian, such
a definition is not appropriate.  

This summarizes the general formulation of the description of the dynamics
of the collective coordinate both at the level of the evolution equation for the expectation value as well as for the effective Langevin dynamics in terms of stochastic noise terms arising from the fluctuations in the meson bath. We are now in condition to study specific models.

\section{TWO MODELS \label{secmodels}}

In the previous sections we established the general aspects of the real-time dynamics of kinks in the presence of the meson bath, obtaining the 
equation of motion as well as the Langevin equation for the collective coordinate
in lowest adiabatic order. Further progress in the understanding of the dynamics
necessarily involves the details of particular models which determine the
matrix elements $T^{(A,S)}$ and therefore the time dependence of the kernels
involved. In this section we study these details for the Sine-Gordon
and $\phi^4$ models. 

\subsection{ Sine-Gordon }

For the Sine-Gordon model the potential is given by 
\begin{equation}
U(\phi) = {m^2\over \lambda} \left( 1 - \cos\left[ {\sqrt \lambda} 
\phi \right] \right)
\label{eigfsg}
\end{equation}
and the static kink solution is given by\cite{d2,c7,c51}
\begin{equation}
\phi_s(x) = {4  \over{\sqrt \lambda}} {\rm arctan} \left[e^{  mx}\right]\,.
\label{phissg}
\end{equation}
 the kink mass and the adiabatic ratio are given by
\begin{equation}
M =  \frac{8m}{\lambda} \; ; \; \frac{m}{M} = \frac{\lambda}{8} \label{massandratio1}
\end{equation}

The normal modes of this theory are the solutions of the following equation, (see eqn.(\ref{hosc})) 
\begin{equation}
\left[ - \frac{d^2}{dx^2} \,+ \,m^2\,-\,\frac{2 m^2}{\cosh^2(mx)}\, \right]\,\psi_n(x)\,=\,\omega^2_n
\psi_n(x). 
\end{equation}
The solutions of the above differential equation are well known \cite{d1},\cite{d2},\cite{c51}. There is only one bound state with zero eigenvalue, the zero mode, followed by a continuum with wave functions given by
\begin{equation}
{\cal U}_k(x) = {1\over{\sqrt{2 \pi}\omega_k}}(-ik \,+ \, m \, {\rm tanh} (m\,x))e^{i k x} 
\label{Usg}
\end{equation}
with $\omega_k^2 = k^2 + m^2$, i.e. $w_0=m$. The scattering states represent the meson  excitations around the kink\cite{jackiwrevmod}.

The matrix elements $D_{pk}$ were already calculated by de Vega\cite{c51}, (see eqns.(\ref{Dmn},\ref{me1})) and are given by
\begin{eqnarray}
D_{kp} \,=\, i p \delta(k + p) + {i ( p^2 - k^2) \over{ 4 \omega_k \omega_p 
\sinh\left[ {\pi \over 2} {(p + k)\over m}\right]}} \quad \hbox{ for }
 p \neq k  
\label{elmatsg}
\end{eqnarray}
which determine the symmetric and antisymmetric matrix elements 
\begin{eqnarray}
T_{pq}^{(S)} &=& {1\over 4} \left[ \left({\omega_p\over{\omega_q}}
\right)^{1/2} - \left({\omega_q\over{\omega_p}}\right)^{1/2}\right] 
\left\{ {( q^2 - p^2) \over{ 4 \omega_k \omega_p 
\sinh\left[ {\pi \over 2} {(q + p)\over m}\right]}} \right\} \nonumber \\
T_{pq}^{(A)} &=& {1\over 4} \left[ \left({\omega_p\over{\omega_q}}
\right)^{1/2} + \left({\omega_q\over{\omega_p}}\right)^{1/2}\right] 
\left\{ {( q^2 - p^2) \over{ 4 \omega_k \omega_p 
\sinh\left[ {\pi \over 2} {(q + p)\over m}\right]}} \right\} \,.
\label{TaTssg}
\end{eqnarray}

Since in this theory there are no bound states other than the zero mode, $F[a^\dagger,a]$ is given only by the first two terms in eqn.(\ref{foobd}). We recognize the ``structure factor''
\begin{equation}
S(Q)= \int_{-\infty}^{\infty} \frac{dx}{2\pi} e^{iQx} \frac{2m^2}{\cosh^2[mx]} = \frac{Q}{\sinh[\frac{Q\pi}{2m}]}
\label{strucfac}
\end{equation}
This structure factor will play an important role in understanding
the large energy behavior of the one-loop contribution. The important
point to notice is that the structure factor is dominated by momenta
$Q \approx m$, falling off exponentially for $|Q| \gg m$.

Substituting eqn.(\ref{TaTssg}) in eqns.(\ref{GTT}) and eqn.({\ref{sigmf}), we obtain the final form of the kernels in this case
\begin{eqnarray}
\Gamma_m(t - t^\prime) &=& {1\over{4^3}}\int_{-\infty}^\infty dQ |S(Q)|^2
 \int_{-\infty}^\infty dk \frac{(Q-2k)^2}{\omega^3_{Q-k} \omega_k^3} 
 \left\{ ( 1 + 2n_k) (\omega_{Q-k} - 
\omega_{k})^2 \sin\left[(\omega_{Q-k} + \omega_{k})
(t - t^\prime) \right] \right.   \nonumber \\
&& \left. \qquad \qquad  -\quad (n_{Q-k}-n_k)  (\omega_{Q-k} + \omega_{k})^2 
\sin\left[(\omega_{Q-k} - \omega_{k})(t - t^\prime)\right] \right\} \label{GTTM}
\end{eqnarray}

\begin{eqnarray}
\Sigma_m(t - t^\prime) &=& {1\over{4^3}}\int_{-\infty}^\infty dQ |S(Q)|^2\, Q
 \int_{-\infty}^\infty dk \frac{(Q-2k)^3}{\omega^3_{Q-k} \omega_k^3} 
 \left\{ ( 1 + n_{Q-k}+n_k) (\omega_{Q-k} - 
\omega_{k}) \; \mbox{\textsf{x}}\right.   \nonumber \\
&& \left.  \cos\left[(\omega_{Q-k} + \omega_{k})
(t - t^\prime) \right] \;  -\; (n_{Q-k}-n_k)  (\omega_{Q-k} + \omega_{k}) 
\cos\left[(\omega_{Q-k} - \omega_{k})(t - t^\prime)\right] \right\} .\nonumber \\
\label{sigmsg}
\end{eqnarray} 

The introduction of $S(Q)$ clarifies that $Q$ is the momentum transferred into the
meson loop, and because $S(Q)$ is peaked at $Q=0$ with a width of the order of the
meson mass we conclude that the  momentum transferred into the meson loop is of the
order of the meson mass. This observation will prove to be very important in the analysis
of the high temperature limit in a later section.

It proves useful to 
express $\Gamma_m(t - t^\prime)$ and $\Sigma_m(t - t^\prime)$ in terms of dimensionless quantities to display at once the nature of the
adiabatic expansion. To achieve this let us make the following change of variables
\begin{equation}
Q  \rightarrow \frac{Q}{m}\quad ;\quad k  \rightarrow \frac{k}{m}\quad ; \quad 
\tau = m\,t\quad \hbox{ and } \quad {\cal T} = {T\over  m} \label{change} \, .
\end{equation}
Then $\Gamma_m(t - t^\prime)$ and $\Sigma_m(t - t^\prime)$ can be written as
\begin{equation}
\Gamma_m(t - t^\prime) = m^2 \Gamma(\tau - \tau^\prime)\quad\quad\hbox{ and } \quad\quad
\Sigma_m(t - t^\prime) = m^3 \Sigma(\tau - \tau^\prime),
\label{dimless}
\end{equation}
where
\begin{equation}
\Gamma(\tau)=\int_{-\infty}^\infty dQ dk |S(Q)|^2
  \left\{\Gamma_1(Q,k)\sin\left[(w_{Q-k} + w_{k})(\tau ) \right]+\Gamma_2(Q,k)\sin\left[(w_{Q-k} - w_{k})(\tau ) \right]\right\} \label{GTTMm} 
\end{equation}
\begin{equation}
\Sigma(\tau)= \int_{-\infty}^\infty dQ dk |S(Q)|^2
 \left\{\Sigma_1(Q,k)\cos\left[(w_{Q-k} + w_{k})(\tau) \right]+\Sigma_2(Q,k)\,\cos\left[(w_{Q-k} - w_{k})(\tau) \right] \right\}
\label{sigsg}
\end{equation}
with
\begin{eqnarray}
\Gamma_1(Q,k) &=& {1\over{64}}\;{( 1 + 2n_k ) (Q-2k)^2 (w_{Q-k}\,-\,w_k)^2 
\over{w^3_{Q-k}\,w_k^3\,}} \nonumber \\
\Gamma_2(p,k) &=&  {1\over{32}}\;{n_k  (Q-2k)^2 (w_{Q-k}\,+\,w_k)^2 
\over{w^3_{Q-k}\,w_k^3\,}} \nonumber \\
\Sigma_1(Q,k) & = & (w_{Q-k}\,+\,w_k)\;\Gamma_1(Q,k) \nonumber \\
\Sigma_2(p,k) & = & (w_{Q-k}\,-\,w_k)\;\Gamma_2(Q,k) \nonumber \\
w_k^2  & = & k^2 +1 \; ; \; 
n_k  =  \frac{1}{e^{\frac{ w_k}{ {\cal T}}}-1}. 
\end{eqnarray}

Fig.(\ref{figssg}) shows the numerical evaluation of $\Gamma(\tau)$ and $\Sigma(\tau)$ vs. $\tau$ for different values of ${\cal T}$. We clearly see
that the self-energy kernel $\Sigma$ is peaked near $\tau =0$ and localized
within a time scale $\tau_s \approx m^{-1} $. Similarly, the kernel $\Gamma$ varies slowly over a large time scale $ \approx 5-10 m^{-1} $.

\subsubsection{Equation of motion:  Exact solution vs. Markovian approximation}

 In terms of dimensionless
quantities the equation of motion (\ref{eqmotm}) becomes in this case

\begin{equation}
\dot{v} (\tau) + \frac{\lambda}{8}\int_0^\tau {\rm d} \tau^\prime \,\Sigma(\tau-\tau^\prime)v(\tau^\prime) = J,
\label{seem}
\end{equation}
where $J = {j}/(mM) \; $and the dot stands for derivative with respect to
the dimensionless variable $\tau$.

As shown in fig.(\ref{figssg}), the kernel $\Sigma(\tau)$  has ``memory'' on time scales a few times the inverse of the meson frequency. 
If the  velocity of the domain wall varies on time scales larger than the ``memory'' of the kernel  a Markovian approximation to the dynamics may be reasonable. The first step in the Markovian approximation corresponds to replacing $v(\tau')$ by $v(\tau)$ inside the integral in eqn. (\ref{seem}) and taking it outside the integral. A second stage of approximation would take the upper limit of the integral to $\infty$ thus integrating the peak of the kernel. However we have shown above that the total integral of the kernel vanishes, thus this second stage cannot be invoked. Recognizing that
$\int_0^{\tau} \Sigma(\tau -\tau')d\tau' = \Gamma(\tau)$ the Markovian 
approximation to (\ref{seem}) is given by

\begin{equation}
\dot{v}(\tau) + \frac{\lambda}{8}v(\tau)\Gamma(\tau)=J.
\label{smem}
\end{equation}

As advanced in the previous section, we now identify the kernel $\Gamma(\tau)$
as the {\em dynamical} friction coefficient in the Markovian approximation.
The property (\ref{zeroint}) determines that $\Gamma(\tau \rightarrow \infty) =0$.

We will now focus on the initial value problem with $J=0$ and $v(\tau=0)=v_0$. The formal solution of the equation of motion in the Markovian approximation is given by
\begin{equation}
v(\tau) = v_0 e^{-\frac{\lambda}{8}\int^{\tau}_0 \Gamma(\tau')d\tau'}
\end{equation}
Even in the Markovian approximation the relaxation of the velocity at
long times is not exponential because  $\Gamma(\tau) \rightarrow 0$ at long times as can be seen in fig. (\ref{figssg}).

\subsubsection{Velocity relaxation and wave function renormalization }

In order to display more clearly the dissipative effects, we now study the
relaxation of the kink velocity. For this consider the initial value problem with $j(t>0) =0$ and initial velocity $v(t=0)=v_0$. 

As the kink moves through the bath, its velocity decreases because of the interaction with the mesons,  the asymptotic final velocity is related to
the initial velocity through the wave function renormalization as explained 
section \ref{gensol} above. We  present the numerical solution of the homogeneous
equation for $v(t)/v_0$ in figure (\ref{figsv}), where we also present the
homogeneous solution in the Markovian approximation described above. We 
clearly see that the initial velocity relaxes to an asymptotic value $v_{\infty}/v_0$. However the time dependence cannot be fit with an exponential. We can see that even at high temperatures the Markovian
approximation grossly fails to describe the dynamics.

According to the analysis of the general solution, 
the ratio $v_{\infty}/ v_0$ should be given by the wave function renormalization, 
i.e.
\begin{equation}
Z_s = \frac{1}{1+\frac{m}{M}\tilde{\Gamma}(s=0)}= \frac{v_{\infty}}{v_0}.\label{wavefc}
\end{equation}

Table 1 below compares the ratio $v_{\infty}/v_0$ obtained from the numerical
solution to the exact evolution equation, with the value of the wave-function renormalization. Clearly the agreement is excellent, confirming the analysis of the asymptotic behavior of the solution in real time.

\begin{center}
\begin{tabular}{||c||c|c||c|c||}
\multicolumn{5}{c}{\textbf{Table (1)}: Numerical evaluation of $Z_s$ and $v_\infty/v_0$ 
in Sine-Gordon theory for $m/M=0.1,0.25$. } \\  
\multicolumn{5}{c}{  } \\ \hline \hline
   & \multicolumn{2}{c||}{ $v_\infty / v_0$ } & \multicolumn{2}{c||}{ $Z_s$ } \\ \hline
   & m/M = 0.1 & m/M = 0.25& m/M = 0.1 & m/M = 0.25 \\  \hline\hline
Zero Temp. &  0.999808 & 0.999521 & 0.999808 & 0.999521 \\ \hline
Temp. 1.0 &  0.993438 & 0.983754 & 0.993438 &  0.983753 \\ \hline
Temp. 5.0 &  0.96055 & 0.906885 &  0.96055 & 0.90687 \\ \hline
Temp. 10.0 & 0.923458 & 0.828352 & 0.923446 & 0.828303 \\ \hline \hline 
\end{tabular}
\end{center}

\subsubsection{ Kernels for the  semiclassical Langevin equation}

Knowledge of the matrix elements $T^{(A)},T^{(S)}$ allow us to obtain
the final form of the kernels that enter in the semiclassical Langevin equation given by eqns.(\ref{ker1}) and (\ref{ker}), and eqn.(\ref{TaTssg}). These kernels can be written in terms of the dimensionless quantities given by eqn.(\ref{change}). 
Since $K_1(t-t') = -2i\Sigma_m(t-t')$ we focus on $K(t-t')$.
In term of dimensionless quantities, $K(t)= m^4 {\cal K}(\tau)$ where

\begin{equation}
{\cal K}(\tau)\,=\, \int_{-\infty}^\infty  dQ d k Q^2 |S(Q)|^2 \left\{C_1(Q,k)\,\cos\left[(w_{Q-k} + w_{k})(\tau) \right]\;+\;C_2(Q,k)\,\cos\left[(w_{Q-k} - w_{k})(\tau) \right] \right\}.
\label{kersingor}
\end{equation}
with
\begin{eqnarray}
C_1(Q,k) &=& {2\over{4^4}}\;{( 1 + n_{Q-k} + n_k + n_{Q-k}\,n_k) (Q-2k)^4  
\over{w^3_{Q-k}\,w_k^3 }} \label{c1singor} \\
C_2(Q,k) &=& {1\over{4^3}}\;{n_k(1 + n_{Q-k}) (Q-2k)^4  
\over{w^3_{Q-k}\,w_k^3 }} \label{c2singor}
\end{eqnarray}
The contribution from $C_1$ is recognized to arise from the process of emission and
annihilation (spontaneous and induced) of two mesons, whereas that from $C_2$ arises
from the scattering off in medium mesons and has its origin in the Landau damping diagram shown in figure (\ref{bubbles}).

Fig.(\ref{figscr}) shows ${\cal K}(\tau)$ for different temperatures ${\cal T}$.
Notice that at large temperatures the kernel becomes strongly peaked at
$\tau =0$ and one would be tempted to conclude that the classical limit
corresponds to a delta function. However the coefficients (\ref{c1singor},\ref{c2singor}) are such that the total integral in $\tau$ (leading to delta functions of sums and differences of frequencies) vanishes. We then conclude that even in the high temperature limit the noise-noise correlation function is not a delta function, i.e. the noise is ``colored'', the classical fluctuation dissipation relation in terms
of a delta function noise correlation does not emerge and  a diffusion coefficient cannot
be appropriately defined. We postpone until a later section a discussion
of the high temperature limit and the classical regime.
 
\subsection{$\phi^4$ Theory}

In this model the potential is given by
\begin{equation}
U(g,\phi) = \frac{m^2}{2\lambda} \left(1 - \lambda \phi^2 \right)^2,
\label{uphi}
\end{equation}
 The static kink solution is given by
\begin{equation}
\phi_s(x-x_0) =  { 1 \over{\sqrt \lambda}} {\rm tanh} \left[m(x-x_0)\right]\,,
\label{phifis}
\end{equation}
and the kink mass is given by
\begin{equation}
M = \frac{4m}{3\lambda} \label{massfi4}
\end{equation}

\noindent and the normal modes are the solutions to the equation, (see eqn.(\ref{hosc}))
\begin{equation}
\left[ - \frac{d^2}{dx^2} \,+ \,4 m^2\,-\,\frac{6 m^2}{{\rm cosh}^2(mx)}\, \right]\,\psi_n(x)\,=\,\omega^2_n
\psi_n(x). 
\end{equation}
The solution of the above differential equation is well known \cite{d1},\cite{d2}. It has two bound states followed by a continuum. The normalized eigenvectors are given by
\begin{eqnarray}
{\cal U}_0(x) &=& \frac{ \sqrt{3\,m} }{2} {\rm sech}^2[m\,x] \propto \frac{d \phi_s}{dx} \quad\quad \mbox{with} \quad\quad
\omega_0\,=\,0 \nonumber \\
{\cal U}_b(x) &=& \frac{ \sqrt{3\,m} }{2} \, {\rm sech}[m\,x] \, {\rm tanh}[m\,x]\quad\quad \mbox{with} \quad\quad \omega_b^2\,=\,3\,m^2 \nonumber \\
{\cal U}_k(x) &=& {m^2 e^{i k x}\over{\sqrt{2\pi (k^2 + m^2)} \omega_k}}
\left\{ 3 {\rm tanh}^2[m x] - 3 \, i \,{k\over m}\, {\rm tanh} [m x] - 1 - \frac{k^2}{m^2} \right\} 
\label{Up4}
\end{eqnarray}
with $\omega_k^2 = k^2 + 4\,m^2$. The scattering states are identified with  meson modes and the  meson frequency is identified with $\omega_o=2m$. 

The bound state with zero frequency is the ``zero mode'', whereas
the bound state with $\omega^2_b = 3m^2$ corresponds to an amplitude distortion\cite{jackiwrevmod,c7} or excited state of the kink. 

The matrix elements $D_{pk}$  are given by, (see eqns.(\ref{Dmn},\ref{me1}))

\begin{eqnarray}
D_{bk} &=& \frac{\sqrt{3 \pi}}{8}\,\frac{ {\rm sech}\left[ \frac{\pi k}{2\,m} \right] }{m^\frac{3}{2}\; \omega_k } \, \sqrt{k^2+m^2}\;(k^2+3\,m^2) \quad\quad \hbox{ (from the bound state) }\nonumber \\
D_{pk} &=& i k \delta(p + k) + {3 \,i \,  \pi ( k^2 - p^2) 
(p^2 + k^2 + 4 m^2)\over{ 4  m^4 N_p\,N_k
\sinh\left[ {\pi \over 2} {(p + k)\over m}\right]}} \quad \hbox{ for }
 p \neq k \; ,  
\label{elmatp4}
\end{eqnarray}
where $N_k$ is defined as
\begin{equation}
N_k = \sqrt{ {2 \pi w_k^2 (k^2 + m^2) \over{m^4}} }\,.
\end{equation}

We notice that the coupling to the continuum through the bound state given by 
the matrix element $D_{bk}$ is of the same order as the coupling to the continuum-continuum (matrix elements $D_{pk}$). This will have interesting consequences  for the dissipational dynamics.  
The symmetric and antisymmetric matrix elements for the continuum states are  given by
\begin{eqnarray}
T_{pq}^{(S)} &=& {3\over 32} \left[ \left({\omega_p\over{\omega_q}}
\right)^{1/2} - \left({\omega_q\over{\omega_p}}\right)^{1/2}\right] 
\left\{ {( q^2 - p^2)(p^2 + q^2 + 4m^2)
\over{ \sqrt{q^2 + m^2} \sqrt{p^2 + m^2} \omega_q \omega_p 
\sinh\left[ {\pi \over 2} {(q + p)\over m}\right]}} \right\} \nonumber \\
T_{pq}^{(A)} &=& {3\over 32} \left[ \left({\omega_p\over{\omega_q}}
\right)^{1/2} + \left({\omega_q\over{\omega_p}}\right)^{1/2}\right] 
 \left\{ {( q^2 - p^2)(p^2 + q^2 + 4m^2)
\over{ \sqrt{q^2 + m^2} \sqrt{p^2 + m^2} \omega_q \omega_p 
\sinh\left[ {\pi \over 2} {(q + p)\over m}\right]}} \right\},
\nonumber \\
&& \mbox{}
\label{TaTsp4}
\end{eqnarray}

\noindent whereas those involving the bound state are obtained by replacing the matrix elements $D_{bk}$ for the $D_{pk}$.

Since in this model there is one bound state other than the zero mode, the interaction vertex $F[a^\dagger,a]$ is given by eqn.(\ref{foobd}) in the 
appendix. 
The contributions from  bound-state-continuum virtual transitions do 
not mix with the continuum-continuum to this order in the adiabatic expansion. As a consequence of this simplification the dimensionless kernels (in terms
of the dimensionless variables introduced in (\ref{change})) become

\begin{eqnarray}
\Gamma(\tau)&=& \int_{-\infty}^\infty {\rm d} p \biggl\{ \; \Gamma_1^b(p)\,\sin\left[(w_p + w_{b})(\tau ) \right]\;+\;\Gamma_2^b(p)\,\sin\left[(w_p - w_{b})(\tau ) \right] \nonumber \\
& & + \quad \int_{-\infty}^\infty dQ |S(Q)|^2
{\rm d} k \left\{\Gamma_1(Q,k)\,\sin\left[(w_{Q-k} + w_{k})(\tau ) \right]\;+\;\Gamma_2(Q,k)\,\sin\left[(w_{Q-k} - w_{k})(\tau )\right]\right\} \;\biggr\}\nonumber \\
\Sigma(\tau)&=&\int_{-\infty}^\infty {\rm d} p \biggl\{ \; \Sigma_1^b(p)\,\cos\left[(w_p + w_{b})(\tau ) \right]\;+\;\Sigma_2^b(p)\,\cos\left[(w_p - w_{b})(\tau ) \right] \nonumber \\
& & + \quad \int_{-\infty}^\infty dQ |S(Q)|^2
{\rm d} k \left\{\Sigma_1(Q,k)\,\cos\left[(w_{Q-k} + w_{k})(\tau) \right]\;+\;\Sigma_2(Q,k)\,\cos\left[(w_{Q-k} - w_{k})(\tau) \right] \right\}
\;\biggr\} \nonumber \\
\label{fisig}
\end{eqnarray}
with
\begin{eqnarray}
\Gamma_1(Q,k) &\equiv& {3^2 \over{4^4}}\;{( 1 + n_{Q-k} + n_k) (Q-2k)^2 (w_{Q-k}\,-\,w_k)^2 \,
((Q-k)^2 + k^2 + 4)^2 \over{w^3_{Q-k}\,w_k^3\,((Q-k)^2 + 1)\, (k^2+1) }} \nonumber \\
\Gamma_2(Q,k) &\equiv&  {3^2 \over{4^4}}\;{(n_k - n_{Q-k}) (Q-2k)^2 (w_{Q-k}\,+\,w_k)^2 \,
((Q-k)^2 + k^2 + 4)^2 \over{w^3_{Q-k}\,w_k^3\,((Q-k)^2 + 1)\, (k^2+1)\,}} \nonumber \\
\Gamma_1^b(p) &\equiv& \frac{\pi \sqrt{3}}{128} \, \frac{(p^4\,+\, 4\,p^2\,+\,3)^2\,(w_p-w_b) (1\,+\,n_b\,+n_p)}{
w_p^3 (w_p+w_b)} {\rm sech}^2 \left[\frac{\pi \,p}{2}\right] \nonumber \\
\Gamma_2^b(p) &\equiv& \frac{\pi \sqrt{3}}{128} \, \frac{(p^4\,+\, 4\,p^2\,+\,3)^2\,(w_p+w_b) (n_b\,-\,n_p)}{
w_p^3 (w_p-w_b)} {\rm sech}^2 \left[\frac{\pi \,p}{2}\right] \nonumber \\
\Sigma_1(Q,k) & \equiv & (w_{Q-k}\,+\,w_k)\;\Gamma_1(Q,k) \; ; \; \Sigma_2(Q,k)  =  
(w_{Q-k}\,-\,w_k)\;\Gamma_2(Q,k) \nonumber \\
\Sigma_1^b(p) & \equiv & (w_p\,+\,w_b)\;\Gamma_1^b(p) \; ; \; \Sigma_2^b(p)  =  (w_p\,-\,w_b)\;\Gamma_2^b(p) \nonumber \\
w_k^2 & = & k^2\,+\,4 \; ,
\end{eqnarray}
where $\Sigma(\tau)$ and $\Gamma(\tau)$ are defined as in eqn.(\ref{dimless}). The functions
$\Sigma(\tau)$ and $\Gamma(\tau)$ where evaluated numerically at different temperatures ${\cal T}$, the results are displayed in fig.(\ref{figpsg}). The behavior of these functions differ from those in the Sine-Gordon theory because of the presence of the bound state which is interpreted as an excited state of the kink.
 As the kink moves in the dissipative medium, energy is transferred between the kink and the bound state resulting in the Rabi-like  oscillations displayed in the figure. We notice that the contribution of the bound state is of the
same order of magnitude as that of the continuum.

\subsubsection{ Equation of motion: Exact solution vs. Markovian approximation}

The solution to the equation of motion and the comparison to the Markovian
approximation proceeds just as in the the case of the Sine-Gordon model.
The equation of motion is again solved as an initial value problem. The exact and Markovian solutions are displayed in figure (\ref{figpv}).

The new feature of the solution are the oscillations that result from virtual
transitions to the bound state. We interpret these in the following manner:
as the kink moves it excites the bound state that corresponds to
a kink distortion\cite{jackiwrevmod}, this excitation in turn reacts-back in the dynamics
of the collective coordinate in a retarded manner. 

 While the exact solution  in this model is qualitatively  similar to that of the Sine-Gordon model, we see however, that quantitatively they are different: there
is stronger dynamical dissipation in the $\phi^4$ model as compared to the Sine Gordon case, due to the strong coupling to the bound state-continuum intermediate states. 

\subsubsection{ Velocity relaxation and wave function renormalization}

In this model the Laplace transform of the functions $\Gamma(\tau)$ and $\Sigma(\tau)$ are given by
\begin{eqnarray}
\tilde{\Sigma}(s) & = & \int_{-\infty}^\infty 
{\rm d} Q \,|S(Q)|^2 {\rm d} k \left\{ 
{\Sigma_1(Q,k)\;s \over{s^2 + (w_{Q-k}\,+\,w_k)^2} } \,+ \,
{\Sigma_2(Q,k)\;s \over{s^2 + (w_{Q-k}\,-\,w_k)^2} } \right\} \nonumber \\
& & \quad + \quad
\int_{-\infty}^\infty 
{\rm d} p  \left\{ 
{\Sigma_1^b(p)\;s \over{s^2 + (w_p\,+\,w_b)^2} } \,+ \,
{\Sigma_2^b(p)\;s \over{s^2 + (w_p\,-\,w_b)^2} } \right\} \nonumber \\
\tilde{\Sigma}(s) & \equiv & s\, \tilde{\Gamma}(s) \nonumber \\
\tilde{\Gamma}(s) & = & \int_{-\infty}^\infty 
{\rm d} Q \,|S(Q)|^2 {\rm d} k \left\{ 
{\Gamma_1(Q,k)\;(w_p\,+\,w_k) \over{s^2 + (w_{Q-k}\,+\,w_k)^2} } \,+ \,
{\Gamma_2(Q,k)\;(w_p\,-\,w_k) \over{s^2 + (w_{Q-k}\,-\,w_k)^2} } \right\} \nonumber \\
& & \quad + \quad
\int_{-\infty}^\infty 
{\rm d} p  \left\{ 
{\Gamma_1^b(p)\;(w_p\,+\,w_b) \over{s^2 + (w_p\,+\,w_b)^2} } \,+ \,
{\Gamma_2^b(p)\;(w_p\,-\,w_b) \over{s^2 + (w_p\,-\,w_b)^2} } \right\}. 
\end{eqnarray}

With the quantities $\Sigma^b \; ; \; \Gamma^b$ given above. 
The homogeneous equations of motion given by (\ref{seem})(exact)  and its Markovian approximation (\ref{smem}) both with $j=0$ are  solved with the kernels $\Sigma(\tau) \; ; \; \Gamma(\tau)$ given above for $v(t)/v_0$. The asymptotic behavior of
the  exact solution will be compared with the prediction $v_{\infty}/v_0 = Z_s$, with
the wave function renormalization $Z_s$ given by eqn.(\ref{wavefc}) but with the $\tilde{\Gamma}(s=0)$ appropriate to the $\phi^4$ model.

Figure (\ref{figpv}) shows the numerical solutions of eqn.(\ref{seem}) and eqn.(\ref{smem}) with $j=0$ for ($v(t)/v_0$) for temperatures ${\cal T}$ = 0, 1.0, 5.0 and 10.0. Again the Rabi-like oscillations associated with
the excitation of the bound state is apparent in the solutions. We have checked numerically that asymptotically the velocity tends to a constant value $v_{\infty}$ but not
exponentially.  Table 2 shows the values of  $v_{\infty}/v_0$ and $Z_s$ for  these temperatures for $m/M\,=\,$0.1 and 0.25 where $v_\infty/v_0$ was evaluated at $\tau = 200$ for the exact solution. Within our numerical errors, we can see that eqn.(\ref{wavefc}) is fulfilled.

\begin{center}
\begin{tabular}{||c||c|c||c|c||} 
\multicolumn{5}{c}{\textbf{Table (2)}: Numerical evaluation of $Z_s$ and $v_\infty/v_0$ 
in $\phi^4$ theory for $m/M=0.1,0.25$. } \\  
\multicolumn{5}{c}{  } \\ \hline \hline
   & \multicolumn{2}{c||}{ $v_\infty / v_0$ } & \multicolumn{2}{c||}
{$Z_s$ } \\ \hline
   & m/M = 0.1 & m/M = 0.25& m/M = 0.1 & m/M = 0.25 \\  \hline\hline
Zero Temp. &  0.999225 & 0.998064 & 0.999176 & 0.997943 \\ \hline
Temp. 1.0 &  0.961561 & 0.911004 & 0.96376 &  0.914072 \\ \hline
Temp. 5.0 &  0.784934 & 0.593058 &  0.787734 & 0.597494 \\ \hline
Temp. 10.0 & 0.642698 & 0.417231 & 0.646577 & 0.422562 \\ \hline \hline 
\end{tabular}
\end{center}

\subsubsection{ Kernels for the semiclassical Langevin equation }

From the definition of the kernels $K_1(t-t^\prime)$ and $K(t-t^\prime)$, eqns.(\ref{ker1}) and (\ref{ker}), and eqn.(\ref{TaTsp4}), these kernels can be written in terms of the dimensionless quantities given by eqn.(\ref{change}) as
\begin{equation}
{\cal K}_1(\tau-\tau^\prime)\,=\, -2\,i\,\Sigma(\tau-\tau^\prime),
\end{equation}
where $\Sigma(\tau-\tau^\prime)$ is given by eqn.(\ref{fisig}) and 
$K(t) = m^4 {\cal K}(\tau)$ with 
\begin{eqnarray}
{\cal K}(\tau) &=& \int_{-\infty}^\infty {\rm d} p \left\{ \; C_1^b(p)\,\cos \left[(w_p + w_{b})(\tau ) \right] \;+\;C_2^b(p)\,
\cos \left[(w_p - w_{b})(\tau ) \right] \right\}  \quad + \nonumber \\
& & \int_{-\infty}^\infty{\rm d} Q |S(Q)|^2 
\int_{-\infty}^\infty{\rm d} k  \left\{ C_1(Q,k)\,\cos\left[(w_{Q-k} + w_{k})(\tau) \right]\;+\;C_2(Q,k)\,\cos\left[(w_{Q-k} - w_{k})(\tau) \right] \right\}\nonumber \\
 \label{calk}
\end{eqnarray}
with the dimensionless matrix elements 
\begin{eqnarray}
C_1(Q,k) &\equiv& {18 \over{4^5}}\;{Q^2( 1 + n_{Q-k} + n_k + n_{Q-k}\,n_k) (Q-2k)^4  ((Q-k)^2 + k^2 + 4)^2 
\over{w_{Q-k}^3\,w_k^3\,((Q-k)^2+1)\,(k^2+1)}} \nonumber \\
C_2(Q,k) &\equiv& {9\over{4^4}}\;{n_k Q^2 (1 + n_{Q-k}) (Q-2k)^4 ((Q-k)^2 + k^2 + 4)^2 
\over{w_{Q-k}^3\,w_k^3\,((Q-k)^2+1)\,(k^2+1)}} \nonumber \\
C_1^b(p) &\equiv& \frac{\pi \sqrt{3}}{4^4} \, \frac{(p^4\,+\, 4\,p^2\,+\,3)^2\,(p^2+1) (1 + n_b + n_p + n_b n_p)}{
w_p^3  {\rm cosh}^2 \left[\frac{\pi \,p}{2}\right]} \nonumber \\
C_2^b(p) &\equiv& \frac{2 \pi \sqrt{3}}{4^4} \, \frac{(p^4\,+\, 4\,p^2\,+\,3)^2\,(p^2+1) n_p(1+\,n_b)}{
w_p^3 {\rm cosh}^2 \left[\frac{\pi \,p}{2}\right]}. 
\end{eqnarray}

Fig.(\ref{figpcr}) shows ${\cal K}(\tau)$ vs. $\tau$  for temperatures ${\cal T}=0,1,5,10$. Again the oscillations
are a consequence of the bound state contribution, and as in the Sine-Gordon case we find that despite the fact that in the high temperature limit the kernel becomes very localized in time, the total integral $\int_{-\infty}^{\infty}d\tau {\cal K}(\tau) =0$ preventing
a representation of the noise-noise correlation function as a delta function in time even in the high temperature limit.  The ``color'' in the noise-noise
correlation function is enhanced by the coupling to the continuum via the bound state which is also responsible for the strong oscillatory behavior of the real-time correlation function.

\section{HARD THERMAL LOOPS VS. CLASSICAL LIMIT \label{secHTloops}}

The high temperature limit corresponds to $T >> m$ with $m$ the meson mass. However we
are restricted to the dilute limit in which the treatment of isolated domain walls is meaningful. Because the kink density is suppressed by an Arrhenius activation factor\cite{habib}
\begin{equation}
N_k \approx e^{-\frac{M}{T}}, \label{kinkdensity}
\end{equation}
the study of the high temperature limit for the dynamics of isolated domain walls requires
that the temperature range be such that
\begin{equation}
m <<  T << M \approx \frac{m}{\lambda}. \label{temprange}
\end{equation}
For weak coupling $\lambda <<1$ there is a wide temperature range in which the
high temperature and the  dilute kink gas approximation will be reliable. In order to understand
the high temperature limit it is convenient to separate the loop integrals
into the ``soft part'' in which both the integrated  and transferred
momenta are ``soft'', i.e. $k,Q << T$ and the ``hard'' part, in which the
loop momentum $k$ is ${\cal O}(T)$. Since the structure factor $S(Q)$ is strongly suppressed for $Q>>m$, the transferred momentum $Q$ is always of order of $m << T$
hence it  is always ``soft''. The ``hard'' $k$ region with
``soft'' transferred momentum is the domain of validity of the hard-thermal
loop resummation programme of Braaten and Pisarski\cite{htl}. 

 Then for the soft region of the remaining
$k$ integral, we can replace the occupation factors $n_k \approx T/\omega_k$. This
soft region therefore gives the {\em classical}  contribution to the kernels $\Sigma, \Gamma$. A simple WKB analysis of the continuum solutions
for both cases considered, reveals that the matrix elements $D_{k,p}$
fall off as $\approx S(p+k)/k$ in the limit in which $Q=k+p \approx m \; ; \; k \rightarrow \infty$. This simple analysis is confirmed by the
exact expression for the matrix elements $T_{k,p}$ (see equations \ref{TaTssg},\ref{TaTsp4}) which in this limit ($Q= k+p \approx m, k \rightarrow \infty$) behave as $S(Q)/k$. Therefore in the meson loop,
the matrix elements yield a contribution of ${\cal O}(1/k^2)$ in the
hard thermal loop limit, for which a simple scaling analysis reveals
a large temperature behavior of ${\cal O}(1/T)$. Hence we see that
in the $1+1$ dimensional case the hard-thermal loop limit yields a
{\em subleading} contribution as compared to the classical contribution
from the soft region. This is a consequence of the small phase space
available for the loop integrals in $1+1$ space-time dimensions. 

This analysis allows us to conclude that the high temperature limit
is dominated by the classical contribution with a lineal dependence
on temperature in the regime $T >> m$. This behavior is clearly 
displayed in figure (\ref{fighit}) that shows the integrals $I_1 \; ; \; I_2$ with
\begin{eqnarray}
I_1 &=& \int_{-\infty}^\infty dk \frac{(Q-2k)^3 (\omega_{Q-k} - 
\omega_{k})}{\omega^3_{Q-k} \omega_k^3} n_k     \nonumber \\
I_2 &=& \int_{-\infty}^\infty dk \frac{(Q-2k)^3 (\omega_{Q-k} + 
\omega_{k})}{\omega^3_{Q-k} \omega_k^3} n_k,
\end{eqnarray} 
corresponding to the contributions from $\Sigma_1 \; ; \; \Sigma_2$ to
the self-energy kernel at $\tau=0$ for Sine Gordon theory, with similar results for
$\phi^4$. We clearly see that for $T \geq 2m$
the temperature dependence becomes lineal. Furthermore we have numerically
checked that most of the contribution in this regime arises from
the ``soft'' region of the loop momentum $k \leq m$ and $Q \leq m$.

Combined together the hard-thermal loop analysis and the numerical evidence lead us to conclude unambiguously that the high temperature
limit of the self-energy kernel is dominated by the classical finite
temperature contribution. 

This analysis also holds for the noise-noise correlation function 
(since the same matrix elements contribute to them). However for the noise-noise correlation function there is an extra factor
of the Bose occupation factors in the integrals. This results in one
extra power of temperature in the ``soft'' region while the temperature 
dependence from the  hard-thermal
loop region is mostly unaffected by the extra Bose factor. Therefore
we conclude that the self-energy kernel is ${\cal O}(m^2 T)$ and the 
noise-noise correlation function is ${\cal O}(m^2 T^2)$ in the high temperature limit. This is in accord with the {\em classical} Fluctuation-Dissipation theorem in which the noise-noise correlation function has
an extra power of temperature compared with the dissipative contribution.

At long times the contribution from the
two meson cut gives a rapidly oscillating contribution leading to a
rapid fall off of the time dependence. On the other hand, the contribution from the Landau damping cut gives the leading contribution at long times
because the discontinuity has support at very low frequencies and
completely dominates long time behavior. Therefore we conclude that
the long time, high temperature behavior in the dilute kink limit is
completely dominated by classical finite temperature dynamics and dominated by the contribution from Landau damping.

\section{HIGHER ORDERS AND HIGHER DIMENSIONS \label{hidim} }

At two-loops and higher orders we expect that collisions will provide a non vanishing static
friction coefficient and result in an exponential relaxation of the velocity in some time regime. However, the contribution from these terms will be of higher order in coupling 
$(m/M)$ and therefore there will be a competition between the time scales associated with 
lowest order relaxation via off-shell Landau damping and the higher order collisional relaxation leading to an exponential fall-off. Therefore we anticipate several different
relaxational regimes with wide separation of the time scales for weak couplings and in the 
dilute regime.

In 3+1 dimensions for degenerate scalar potentials the situation is clearly more complicated.
The zero mode from translation invariance now gives rise to two-dimensional massless degree
of freedom corresponding to small local distortions perpendicular to the (planar) wall. These
are the capillary waves fluctuations of the interface that will dominate the long-wavelength
small frequency dynamics. We expect to report on further studies of higher order collisional
relaxation as well as new phenomena in 3+1 dimensions in the near future.

\section{ CONCLUSIONS AND FURTHER QUESTIONS \label{seccon}}

We have studied the non-equilibrium dynamics of domain walls 
in $1+1$ dimensional scalar field theories at finite temperature in
the dilute regime. We obtained the real time equations of motion for the expectation value of the collective coordinate and also the quantum Langevin equation to lowest order in the weak coupling (adiabatic) expansion. Two specific models were studied: $\phi^4$ and Sine-Gordon scalar field theories providing detailed
analytic and numerical studies of the equations of motion and a Markovian
approximation to it.

To lowest order in weak coupling we found that the real-time
equation of motion involves a non-Markovian self-energy kernel and that the
static friction coefficient vanishes. However, there is dynamical friction
which is a result of the memory effects in the
self-energy and is associated with two different types of two-meson processes: spontaneous and induced two-meson creation and annihilation and
scattering off in medium mesons.
The second type processes only occur at finite temperature and lead to
Landau damping. 

 We studied the Markovian approximation and shown numerically that this approximation is unreliable in a wide range of temperatures. 

The quantum Langevin equation was obtained by computing the influence functional obtained by tracing out the meson  degrees of freedom to the same order in the adiabatic expansion. We found that the dissipative kernel and the noise 
correlation function obey a generalized form of fluctuation-dissipation relation but that a Markovian limit is not available, the noise is Gaussian, additive but colored. 

The high temperature limit in the dilute regime was studied in detail by
analyzing the ``soft'' and ``hard'' contributions to the self energy 
and noise kernels. We find that the hard contribution is suppressed at
high temperature because the matrix elements fall of as an inverse
power of the hard momentum. The small one dimensional phase space leads
to a suppression of the hard momenta and therefore the leading contribution at high temperature arises from the ``soft'' region with
momenta of the order of the meson mass yielding the classical result in
the high temperature limit. Furthermore the long time dynamics is completely determined by the Landau damping processes in the medium leading to the conclusion that to the order studied the long time dynamics is completely determined by {\em classical} Landau damping. 

We have restricted our study to a perturbative expansion which already
showed the complicated nature of the problem even at lowest order. Pursuing a higher order calculation in perturbation theory will clearly
be a major task. When the velocity of the soliton becomes very large
one must abandon the approach advocated in this article and pursue a
non-perturbative approach that accounts for strongly non-linear processes.
We are currently implementing such an approach in terms of a self-consistent variational method\cite{cooperetal} and expect to report on
new results in the near future.

\section*{Acknowledgments}
The authors would like to thank  D. Jasnow, H. de Vega, J. Levy, R. Willey and D. Campbell for helpful discussions and comments. D. B. thanks S. Khlebnikov and P. Arnold for bringing 
Refs.\cite{khleb,arnold} to his attention and for discussions. The authors also thank N.S.F. for partial support through grant awards: PHY-9605186, INT-9512798 and the Pittsburgh Supercomputing Center for grant: PHY-950011P. S. M. A. thanks King Fahad University of Petroleum and Minerals (Saudi Arabia) for
financial support. F. I. T. thanks the Dept. of Physics, Univ. of Pittsburgh for hospitality and CNPq and FAPEMIG for financial support.   




\appendix

\section{REAL-TIME MESON CORRELATION FUNCTIONS \label{A}}

In this appendix, we will calculate the Green's functions which are defined in eqn.(\ref{greens}) in terms of the vertex given by
eqn. (\ref{Faa}). 

Applying Wick's theorem and eqn.(\ref{propaa}), it is a matter of
 straightforward algebra to find the following results:
\begin{eqnarray}
G^{++}(t,t^\prime) & = &  -2 \,\sum_{p,k\not=0} \Biggl\{ T_{pk}^{(S)} \,
T_{-p-k}^{(S)} \biggl[ \,e^{-i (\omega_p + \omega_k)(t - t^\prime) } \Bigl( n_p n_k + \theta(t - t^\prime)( 1 +  n_p + n_k ) \Bigr) \nonumber \\
& & \hspace{1.2in} + \quad e^{+ i (\omega_p + \omega_k)(t - t^\prime) } \Bigl( n_p n_k + \theta(t^\prime - t)( 1 +  n_p + n_k ) \Bigr) \biggr] \nonumber \\
& & \quad \;  + \;\; 2\, T_{pk}^{(A)} \,
T_{-p-k}^{(A)} \biggl[e^{- i (\omega_p - \omega_k)(t - t^\prime) } \Bigl( n_p n_k + n_p\, \theta(t^\prime - t) +  n_k \, \theta(t - t^\prime) \Bigr) \biggr] \Biggr\}
\nonumber \\
& & \nonumber \\
& = &  G^>(t,t')\theta(t-t')+G^<(t,t')\theta(t'-t) \nonumber\\
& & \nonumber \\
G^{--}(t,t^\prime) & = &  -2 \,\sum_{p,k\not=0} \Biggl\{ T_{pk}^{(S)} \,
T_{-p-k}^{(S)} \biggl[ \,e^{-i (\omega_p + \omega_k)(t - t^\prime) } \Bigl( n_p n_k + \theta(t^\prime - t)( 1 +  n_p + n_k ) \Bigr) \nonumber \\
& & \hspace{1.2in} + \quad e^{+ i (\omega_p + \omega_k)(t - t^\prime) } \Bigl( n_p n_k + \theta(t - t^\prime)( 1 +  n_p + n_k ) \Bigr) \biggr] \nonumber \\
& & \quad \;  + \;\; 2\, T_{pk}^{(A)} \,
T_{-p-k}^{(A)} \biggl[e^{- i (\omega_p - \omega_k)(t - t^\prime) } \Bigl( n_p n_k + n_k\, \theta(t^\prime - t) +  n_p \, \theta(t - t^\prime) \Bigr) \biggr] \Biggr\}
\nonumber \\
& & \nonumber \\
& = &  G^>(t,t')\theta(t'-t) + G^<(t,t')\theta(t-t') \nonumber \\
& & \nonumber \\
G^{+-}(t,t^\prime) & = &  2 \,\sum_{p,k\not=0} \Biggl\{ T_{pk}^{(S)} \,
T_{-p-k}^{(S)} \biggl[ \,e^{-i (\omega_p + \omega_k)(t - t^\prime) } \; n_p n_k \, + \, e^{+ i (\omega_p + \omega_k)(t - t^\prime) }\Bigl( n_p n_k + n_p + n_k + 1) \Bigr) \biggr]\nonumber \\
& & \quad \;  + \;\; 2\, T_{pk}^{(A)} \,
T_{-p-k}^{(A)} \biggl[e^{- i (\omega_p - \omega_k)(t - t^\prime) } \; n_p \Bigl(1 +  n_k \Bigr) \biggr] \Biggr\} \nonumber \\
& & \nonumber \\
& = &  -G^<(t,t') \nonumber \\
& & \nonumber \\
G^{-+}(t,t^\prime) & = & 2 \,\sum_{p,k\not=0} \Biggl\{ T_{pk}^{(S)} \,
T_{-p-k}^{(S)} \biggl[ \,e^{-i (\omega_p + \omega_k)(t - t^\prime) } \Bigl( n_p n_k + n_p + n_k + 1) \, + \, e^{+ i (\omega_p + \omega_k)(t - t^\prime) } \; n_p n_k  \biggr]\nonumber \\
& & \quad \;  + \;\; 2\, T_{pk}^{(A)} \,
T_{-p-k}^{(A)} \biggl[e^{- i (\omega_p - \omega_k)(t - t^\prime) } \; n_k \Bigl(1 +  n_p \Bigr) \biggr] \Biggr\}\nonumber \\
& = &  -G^>(t,t')= -G^<(t',t).  
\label{GREENS}
\end{eqnarray}

These Green's functions satisfy the following relation
\begin{equation}
G^{++} \, + \, G^{--} \, + \,G^{+-} \, + \,G^{-+} \, = \,0
\end{equation}
which is a consequence of unitary time evolution\cite{b6}.

Furthermore, using the antisymmetry property of the matrix elements
$T^{(A)}_{pk}$ one finds that 
\begin{equation}
G^{+-}(t,t') = (G^{-+}(t,t'))^* \label{conju}
\end{equation}
The Green's functions $G^{++}(t,t')\; ; \; G^{--}(t,t')$ can be
written in terms of $G^{+-}(t,t')$ and its complex conjugate, therefore
we see that there is only one independent Green's functions (and its
complex conjugate). 

\section{CALCULATING $K_1(t-t^\prime)$ AND $K(t-t^\prime)$ \label{B} }

Performing the coordinate transformation in eqn.(\ref{wigner}), the influence-functional becomes
\begin{eqnarray}
{\cal F}[\dot{x},\dot{R}] & = & exp\left\{-\, \frac{1}{2} \int dt\, dt^\prime \right. \biggl[ \frac{\dot{R}(t) \,\dot{R}(t^\prime)}{4} \biggl(  G^{++}(t,t^\prime) \,+\, G^{--}(t,t^\prime) \,-\, G^{+-}(t,t^\prime)\,-\, G^{-+}(t,t^\prime) \biggr) \nonumber \\ 
& + &  \biggl\{
\frac{1}{2} \dot{R}(t) \,\dot{x}(t^\prime) \biggl(  G^{++}(t,t^\prime) \,-\, G^{--}(t,t^\prime) \,+\, G^{+-}(t,t^\prime)\,-\, G^{-+}(t,t^\prime) \biggr) \nonumber \\
& + &  \left. \;
\frac{1}{2} \dot{x}(t) \,\dot{R}(t^\prime) \biggl(  G^{++}(t,t^\prime) \,-\, G^{--}(t,t^\prime) \,-\, G^{+-}(t,t^\prime)\,+\, G^{-+}(t,t^\prime) \biggr)
\biggr\} \biggr] \right\}.
\end{eqnarray}

Integrating the linear term in $\dot{R}$ by parts once and the quadratic term twice, the influence-functional can be cast in the following form
\begin{equation}
{\cal F}[\dot{x},\dot{R}]  =  exp\left\{\, \frac{1}{2} \int dt\, dt^\prime \Bigl[
R(t) \, K_1(t-t^\prime) \,\dot{x}(t^\prime) \,-\, R(t) \,K(t-t^\prime)\, \dot{R}(t^\prime) \Bigr] \right\},
\end{equation}
where
\begin{eqnarray}
K_1(t-t^\prime) & = & \frac{1}{2} \frac{\partial}{\partial \, t} \biggl[\Bigl(  G^{++}(t,t^\prime) \,-\, G^{--}(t,t^\prime) \,+\, G^{+-}(t,t^\prime)\,-\, G^{-+}(t,t^\prime) \Bigr) \nonumber \\
& & + \quad \;\; \Bigl(  G^{++}(t^\prime,t) \,-\, G^{--}(t^\prime,t) \,-\, G^{+-}(t^\prime,t)\,+\, G^{-+}(t^\prime,t) \Bigr) \biggr] \nonumber\\
& = & 2 \frac{\partial}{\partial \, t}\left[G^>(t,t')-G^<(t,t')\right]\theta(t-t') \\ 
K(t-t^\prime) & = & \frac{1}{4} \frac{\partial^2}{\partial \, t^2} \biggl[ G^{++}(t,t^\prime) \,+\, G^{--}(t,t^\prime) \,-\, G^{+-}(t,t^\prime)\,-\, G^{-+}(t,t^\prime) \biggr]\nonumber \\
& = & \frac{1}{2} \frac{\partial^2}{\partial \, t^2}
\left[G^>(t,t')+G^<(t,t')\right].
\end{eqnarray}

Substituting the values of the Green's functions from eqn.(\ref{GREENS}) in the above equations, one obtains the expressions for $K_1(t-t^\prime)$ and $K(t-t^\prime)$ in eqns.(\ref{ker1}) and (\ref{ker}).

In the case that there are bound states other than the zero mode, such as the case of $\phi^4$ the sum in eqn.(\ref{Faa}) runs over all bound and scattering states, i.e.
\begin{eqnarray}
F[a^\dagger, a] &=&  \frac{1}{2\,i} \int dp\,dk \sqrt{ {\omega_p 
\over{\omega_k}}} D_{pk} \left[ a_k a_p - a_{-k}^\dagger 
a_{-p}^\dagger + a_{-k}^\dagger a_p - a_{-p}^\dagger a_k \right] \nonumber \\
& & + \;\frac{1}{2\,i} \sum_b \int dk \sqrt{ {\omega_b 
\over{\omega_k}}} D_{bk} \left[ a_k a_b - a_{-k}^\dagger 
a_{b}^\dagger + a_{-k}^\dagger a_b - a_{b}^\dagger a_k \right] \nonumber \\
& & + \; \frac{1}{2\,i} \sum_b \int dk \sqrt{ {\omega_k 
\over{\omega_b}}} D_{kb} \left[ a_b a_k - a_{b}^\dagger 
a_{-k}^\dagger + a_{b}^\dagger a_k - a_{-k}^\dagger a_b \right] \nonumber \\
& & + \; \frac{1}{2\,i} \sum_{a,b} \sqrt{ {\omega_a 
\over{\omega_b}}} D_{ab} \left[ a_b a_a - a_{b}^\dagger 
a_{a}^\dagger + a_{b}^\dagger a_a - a_{a}^\dagger a_b \right], \label{lll}
\end{eqnarray}
where the indices $a$ and $b$ stand for summation over discrete bound states and $p$ and $k$ stand for summation over continuum scattering states. The models which we considered in this paper have at most one bound state, that is the case in the $\phi^4$ theory. In this case, the last term will not contribute since $D_{bb}$ vanishes. Thus for only one bound state, eqn.(\ref{lll}) can be written as
\begin{eqnarray}
F[a^\dagger, a] &=& \int dp\,dk\, \left[T^{(S)}_{pk}\,\left(a_p\,a_k \,-\,a^\dagger_{-p}\,
a^\dagger_{-k}\right)\,+\,T^{(A)}_{pk}\,\left(a^\dagger_{-p}\,a_k \,-\,a^\dagger_{-k}\,
a_{p}\right) \right]\nonumber \\
& & + \;
\int dk\, \left[T^{(S)}_{bk}\,\left(a_k\,a_b \,-\,a^\dagger_{-k}\,
a^\dagger_{b}\right)\,+\,T^{(A)}_{bk}\,\left(a^\dagger_{-k}\,a_b \,-\,a^\dagger_{b}\,
a_{k}\right) \right], \label{foobd}
\end{eqnarray}
where the matrices $T^{(S)}_{pk}$ and $T^{(A)}_{pk}$ for scattering states are given by eqn.(\ref{scm})  
and if one of the states is a bound state, then
\begin{eqnarray}
T_{bk}^{(S)} &=& {1\over{2 i}} \left[\sqrt{{\omega_b 
\over{\omega_k}}} - \sqrt{{\omega_k\over{\omega_b}}}\, 
\right] D_{bk} \nonumber \\
T_{bk}^{(A)} &=& {1\over{2 i}} \left[\sqrt{{\omega_b
\over{\omega_k}}} + \sqrt{{\omega_k\over{\omega_b}}}\, 
\right] D_{bk} \,.
\label{Tsa}
\end{eqnarray}

In the Sine-Gordon theory, the last two terms in eqn.(\ref{foobd}) do not contribute since in this theory there are no bound states other than the zero mode and the Green's functions are given by eqn.(\ref{GREENS}) but with integration over $p$ and $k$ instead of the summation.

In the $\phi^4$ case, to lowest adiabatic order the contributions from the bound and scattering states decouple. This implies that  the Green's functions will have a contribution from the bound state which is given by the same expression as that of the scattering states, with $p \rightarrow b$, but multiplied by a factor of $1/2$ since the bound state wave function is chosen to be real.

\section{GENERALIZED FLUCTUATION-DISSIPATION RELATION\label{C} }

The functions
\begin{eqnarray}
G^>(t-t') & = & \langle F(t) F(t') \rangle \\
G^<(t-t') & = & \langle F(t') F(t) \rangle, 
\end{eqnarray}
(where $F$ is given by eqn.(\ref{Faa})) admit a spectral representation,
and their Fourier transforms in time, $g^>(\omega)\; ; \; g^<(\omega)$
 obey the KMS condition\cite{fetter}
\begin{equation}
g^<(\omega) = e^{-\beta \omega} g^>(\omega). 
\end{equation}
From this expression we find that $k_1(\omega)$, the Fourier transform in time of the kernel $iK_1(t-t')= 2 \Sigma_m(t-t')$ is given by
\begin{equation}
k_1(\omega) = 2 \int \frac{d\omega'}{2\pi} \frac{\omega' g^>(\omega')\left[1-e^{-\beta \omega'}\right]}{\omega-\omega'+i\epsilon},
\end{equation}
leading to the imaginary part
\begin{equation}
Im[k_1(\omega)] = -\omega  g^>(\omega)\left[1-e^{-\beta \omega}\right].
\end{equation}

On the other hand the kernel that determines the noise-noise correlation
function $K(t-t')$ has a Fourier transform given by $k(\omega)$ with
\begin{eqnarray}
k(\omega) & = & -\frac{\omega^2}{2}\left[g^>(\omega)+g^<(\omega)\right] = 
-\frac{\omega^2 }{2} g^>(\omega)\left[1+e^{-\beta \omega}\right] \nonumber\\
& = & \frac{\omega}{2} Im[k_1(\omega)]\coth\left[\frac{\beta \omega}{2}\right].
\label{flucdiss}
\end{eqnarray}
The relation between the Fourier transform of the noise-noise correlation
function and the imaginary part of the self-energy is the generalized Fluctuation-Dissipation relation\cite{d8}.


\newpage

%
%

\begin{center}
\begin{huge}
\textbf{Figure Captions}
\end{huge}
\end{center}

\textbf{Figure 1}
The non-equilibrium one-loop contributions to the self energy. The upper
two contributions correspond to emission-annihilation of two mesons. The lower two correspond  scattering off in-medium mesons and responsible for
Landau damping.   \label{bubbles}

\textbf{Figure 2}
Contour in the complex s-plane for the inverse Laplace transform.  \label{cut}


\textbf{Figure 3}
The functions $\Gamma(\tau)$ and $\Sigma(\tau)$ for temperatures ${\cal T}=$ 0, 1.0, 5.0 and 10.0 for Sine-Gordon theory. \label{figssg}


\textbf{Figure 4}
Numerical evaluation of the velocity of the kink for $j=0 \; ; \; v_0=1$ for temperatures ${\cal T}=$ 0, 1.0, 5.0 and 10.0 in Sine-Gordon theory. \label{figsv}



\textbf{Figure 5}
The correlation function ${\cal K}(\tau)$ for temperatures ${\cal T}=$ 0, 1.0, 5.0 and 10.0 in the Sine-Gordon theory. \label{figscr}



\textbf{Figure 6}
The functions $\Gamma(\tau)$ and $\Sigma(\tau)$ for temperatures ${\cal T}=$ 0, 1.0, 5.0 and 10.0 in the $\phi^4$-theory. Contributions from
bound and scattering states are displayed separately.  \label{figpsg}


\textbf{Figure 7}
Numerical evaluation of the velocity of the kink for $j=0\; ; \; v_0=1$ for temperatures ${\cal T}=$ 0, 1.0, 5.0 and 10.0 in $\phi^4$ theory.\label{figpv}



\textbf{Figure 8}
The correlation function ${\cal K}(\tau)$ for temperatures ${\cal T}=$ 0, 1.0, 5.0 and 10.0 in the $\phi^4$ theory. \label{figpcr}



\textbf{Figure 9}
Integrals $I_1 \; ; I_2$ corresponding to the contributions from $\Sigma_1 \; ; \; \Sigma_2$ to
the self-energy kernel at $\tau=0$  vs. ${\cal T}$  in the Sine Gordon theory. \label{fighit}


\end{document}